\documentclass[11pt,a4paper]{article}
\usepackage{jheppub}
\pdfoutput=1

\usepackage{amsmath, amsfonts, amssymb, slashed, bm, bbold}
\usepackage{color, cleveref, graphicx, setspace}
\usepackage{subcaption, float}
\usepackage{physics, cancel}

\usepackage[english]{babel}
\usepackage[utf8]{inputenc}
\usepackage[T1]{fontenc}

\usepackage{ifthen}

\newcommand{\y}[2][]{   
    \ifthenelse{\equal{#1}{} }{ \tilde{y}^{(5)}_{#2} }{ \tilde{y}^{(5)}_{#2,\,#1} } 
}
\newcommand{\yC}[2][]{   
    \ifthenelse{\equal{#1}{} }{ \tilde{y}^{(5)\,*}_{#2} }{ \tilde{y}^{(5)\,*}_{#2,\,#1} } 
}
\newcommand{\my}[2][]{   
    \bm{[} \tilde{y}^{(5)}_{#2} \bm{]}_{#1}
}
\newcommand{\myC}[2][]{   
    \bm{[} \tilde{y}^{(5)}_{#2} \bm{]}^{*}_{#1}
}

\renewcommand{\[}{\left[}
\renewcommand{\]}{\right]}
\renewcommand{\(}{\left(}
\renewcommand{\)}{\right)}
\def\be{\begin{equation}}
\def\ee{\end{equation}}
\def\bes{\begin{equation*}}
\def\ees{\end{equation*}}
\def\bead{\begin{aligned}}
\def\eead{\end{aligned}}
\def\bmat{\left(\begin{matrix}}
\def\emat{\end{matrix}\right)}

\def\cL{{\cal L}}

\begin{document}

\title{Flavoured Warped Axion}

\preprint{\begin{tabular}{r} UMN--TH--4007/20\\ DESY 20-218\\ CPHT-RR094.122020 \end{tabular}}

\author[a]{Quentin Bonnefoy,} 
\author[b]{Peter Cox,} 
\author[c]{Emilian Dudas,} 
\author[d]{Tony Gherghetta,}
\author[d]{Minh D. Nguyen}

\affiliation[a]{Deutsches Elektronen-Synchrotron DESY, 22607 Hamburg, Germany}
\affiliation[b]{School of Physics, The University of Melbourne, Victoria 3010, Australia}
\affiliation[c]{CPHT Ecole Polytechnique, CNRS, Institute Polytechnique de Paris, 91128 Palaiseau, France}
\affiliation[d]{School of Physics and Astronomy, University of Minnesota, Minneapolis, Minnesota 55455, USA}

\emailAdd{quentin.bonnefoy@desy.de}
\emailAdd{peter.cox@unimelb.edu.au}
\emailAdd{emilian.dudas@polytechnique.edu}
\emailAdd{tgher@umn.edu}
\emailAdd{nguy1642@umn.edu}

\abstract{We consider a 5D extension of the DFSZ axion model that addresses both the axion quality and fermion mass hierarchy problems, and predicts flavour-dependent, off-diagonal axion-fermion couplings. The axion is part of a 5D complex scalar field charged under a U(1)$_{\rm PQ}$ symmetry that is spontaneously broken in the bulk, and is insensitive to explicit PQ breaking on the UV boundary. Bulk Standard Model fermions interact with two Higgs doublets that can be localized on the UV boundary or propagate in the bulk to explain the fermion masses and mixings. When the Higgs doublets are localized on the UV boundary, they induce flavour diagonal couplings between the fermions and the axion. 
However, when the Higgs doublets propagate in the bulk, the overlap of the axion and fermion profiles generates flavour off-diagonal couplings. The effective scale of these off-diagonal couplings in both the quark and lepton sectors can be as small as $10^{11}$\,GeV, and therefore will be probed in future precision flavour experiments.}

\maketitle


\section{Introduction}

A popular solution of the strong CP problem involves extending the Standard Model by introducing a new particle, the axion. This pseudoscalar boson arises from the Peccei-Quinn (PQ) mechanism~\cite{Peccei:1977hh}, where an anomalous U(1)$_{\rm PQ}$ global symmetry is spontaneously broken~\cite{Weinberg:1977ma,Wilczek:1977pj}. When nonperturbative QCD corrections generate a potential, the axion relaxes to its CP-conserving minimum, solving the strong CP problem. Furthermore, for some mass ranges, the axion can be a good cold dark matter candidate~\cite{Preskill:1982cy,Abbott:1982af,Dine:1982ah}. The fact that two problems in the Standard Model are simultaneously addressed has led to an extensive experimental effort in searching for the axion. 

An underlying assumption of the axion solution is that the U(1)$_{\rm PQ}$ global symmetry is well-preserved. However, this is in conflict with the expectation that global symmetries are explicitly broken by gravity~\cite{Banks:2010zn,Banks:1988yz}. In order not to misalign the minimum of the axion potential and reintroduce the strong CP problem, the global symmetry must then be preserved up to high-dimension terms in the effective Lagrangian~\cite{Holman:1992us,Kamionkowski:1992mf, Barr:1992qq, Ghigna:1992iv}. Using a slice of AdS$_5$~\cite{Randall:1999ee}, the axion-quality problem was recently addressed~\cite{Cox:2019rro} by delocalizing a bulk axion zero mode away from the UV brane, which sources explicit (gravitational) violations of the PQ symmetry. By the AdS/CFT correspondence~\cite{Maldacena:1997re}, this 5D geometric solution has a dual 4D interpretation where the U(1)$_{\rm PQ}$ symmetry is an accidental global symmetry of some underlying strong dynamics, such as that considered in \cite{Gavela:2018paw}.

The advantage of solving the axion-quality problem in a slice of AdS$_5$ is that one can use the 5D geometry to also explain the Standard Model fermion mass hierarchy. 
In this work, we extend the model considered in Ref.~\cite{Cox:2019rro}, containing a PQ-charged, bulk complex scalar field, to also include bulk Standard Model fermions. The Higgs sector, consisting of two Higgs doublets, can either be localized on the UV boundary or propagate in the bulk. This 5D model is essentially the DFSZ model~\cite{Zhitnitsky:1980tq,Dine:1981rt} with the singlet scalar field, containing the axion, propagating in the bulk. The fact that the PQ symmetry is gauged in the bulk, is also in agreement with the expectation that only gauge symmetries are present in quantum gravity.

A feature of this 5D model is that the axion-fermion couplings are obtained while automatically addressing the fermion mass hierarchy and axion quality problem, unlike the original 4D DFSZ model. The bulk fermion profiles are controlled by order one 5D fermion mass parameters. Once these parameters are chosen to explain the Standard Model fermion mass hierarchy and mixings, they give predictions for the axion couplings to fermions. For a boundary-localized Higgs sector, only flavour-diagonal couplings are generated. This follows from the orthonormality of the bulk fermion profiles. However, when the Higgs sector propagates in the bulk, there is a non-trivial wavefunction overlap between the axion and the fermion profiles that gives off-diagonal fermion couplings. 
The predictions for the off-diagonal couplings involving quarks and charged leptons are consistent with the current experimental limits~\cite{MartinCamalich:2020dfe,Calibbi:2020jvd}.
Assuming an axion decay constant $F_a \sim 10^9$\,GeV, the effective scale of the axion-fermion off-diagonal couplings is of order $10^{11}-10^{15}$\,GeV.

Furthermore, we also discuss the axion couplings to gluons and photons. In particular, using the known form of the 5D anomaly~\cite{ArkaniHamed:2001is,Hirayama:2003kk}, we derive the axion-gluon/photon couplings. These couplings can also be directly calculated from a Kaluza-Klein sum over 4D fermion modes, which provides a nontrivial check of our results. A 5D Chern-Simons term can also generate an axion coupling to gauge bosons, and we show how such interactions arise from integrating out bulk fermions, extending the calculation of Ref.~\cite{Witten:1996qb}.

The outline of our paper is as follows. In Section \ref{sec:5Daxionmodel} we review the 5D axion model~\cite{Cox:2019rro}, and then introduce bulk Standard Model fermion fields. There are two choices for the Higgs fields. UV boundary Higgs are first considered in Section \ref{sec:boundaryHiggs} where the axion-fermion couplings are shown to be flavour diagonal. Next, in Section \ref{bulkHiggsSection}, we consider the bulk Higgs case and derive the flavour-dependent, off-diagonal axion-fermion couplings for both a massless and massive axion in Section~\ref{sec:axionfermionBulk}. The axion-gluon/photon couplings from both the 5D anomaly and Chern-Simons term are discussed in Section~\ref{sec:axion-gluonphoton}. The concluding remarks are presented in Section~\ref{sec:conclusion}.  The appendices contain further details of our calculations. In appendix~\ref{app:SMflavour} we present the approximations used in obtaining the Standard Model fermion masses and mixings for both the quark and lepton sector. Appendix~\ref{KKcomputationAppendix} contains a direct 4D calculation of the axion couplings to gauge bosons that uses the axion interactions with the fermionic KK modes. The boundary axion couplings can be related to a 5D Chern Simons interaction and this connection is presented in appendix~\ref{CStermAppendix}.


\section{The 5D Axion Model}
\label{sec:5Daxionmodel}

Consider a 5D $U(1)_{PQ}$ gauge theory in a slice of AdS$_5$. The metric is given by
\begin{equation}
  ds^2=A^2(z)\left(dx^2+dz^2\right)\equiv g_{MN} dx^M dx^N\,,
  \label{eq:AdSmetric}
\end{equation}
with coordinates $x^M=(x^\mu,z)$, and where $A(z)=1/(kz)$ with $k$ the AdS curvature scale. We denote the $U(1)_{PQ}$ gauge field by $V_M=(V_\mu,V_z)$ and introduce a complex scalar $\Phi=\eta\, e^{ia}$ with $PQ$ charge $X_\Phi=1$. The action is~\cite{Cox:2019rro}
\begin{align} \label{eq:5D_action}
  S =~& 2\int^{z_{IR}}_{z_{UV}}d^5x\, \sqrt{-g} \left( -\frac{1}{4g_5^2}F^{MN}F_{MN} - \frac{1}{2} \big(\mathcal{D}^M\Phi\big)^\dagger\big(\mathcal{D}_M\Phi\big) - \frac{1}{2}m_\Phi^2\Phi^\dagger\Phi \right. \notag \\
  &\left.-\frac{1}{2g_5^2\xi_{PQ}}\left( g^{\mu\nu}\partial_\mu V_\nu + \xi_{PQ} A^{-3}\partial_z\left(AV_z\right) - \xi_{PQ} g_5^2 X_\Phi \eta^2 a \right)^2 \right) \notag \\
  &- \int d^4x\, \sqrt{-g_4} \, U(\Phi) \,,
\end{align}
where $\mathcal{D}_M=\partial_M -iX_\Phi V_M$, $g_5$ is the 5D gauge coupling and $\xi_{PQ}$ is a gauge-fixing  parameter.
The scalar potentials on the UV and IR branes, located at $z=z_{UV}$ and $z=z_{IR}$ respectively, are taken to be
\begin{align}
  U_{UV}(\Phi) &=  b_{UV} k\, \Phi^\dagger\Phi \,, \label{eq:UV-potential} \\
  U_{IR}(\Phi) &= \frac{\lambda_{IR}}{k^2}\left(\Phi^\dagger\Phi-k^3v_{IR}^2\right)^2 \,. \label{eq:IR-potential}
\end{align}
Neglecting the backreaction of the scalar field on the metric\footnote{This requires $|(\partial_z\eta)^2-m_\Phi^2\eta^2|\ll 12k^2M_5^3$, where the 5D Planck mass $M_5$ is related to the 4D Planck mass via $M_P^2 \simeq M_5^3/k$.}, the equation of motion for the scalar yields the background solution
\begin{equation} \label{eq:bulk-scalar}
  \eta(z) = k^{3/2} \left( \lambda\, (kz)^{4-\Delta} + \sigma\, (kz)^{\Delta} \right) \,,
\end{equation}
where $\Delta>2$ is related to the bulk scalar mass via $m_\Phi^2=\Delta(\Delta-4)\,k^2$.
The real parameters $\sigma$ and $\lambda$ are determined by the boundary conditions: 
\begin{align}
  \sigma &= \sqrt{v_{IR}^2-\frac{\Delta}{2\lambda_{IR}}}\,(kz_{IR})^{-\Delta}\equiv \sigma_0\,(kz_{IR})^{-\Delta} \,, \\
  \lambda &= \frac{\Delta-b_{UV}}{\Delta-4+b_{UV}}(kz_{UV})^{2\Delta-4}\sigma \,,
\end{align}
assuming $z_{UV} \gg z_{IR}$.


\subsection{Axion profile}

The action \eqref{eq:5D_action} leads to coupled equations of motion for the scalar degrees of freedom, $a(x^\mu,z)$ and $V_z(x^\mu,z)$.  
These can be solved via the Kaluza-Klein (KK) expansion:
\begin{align} \label{eq:KK-expansion}
  a(x^\mu,z) &= \sum_{n=0}^\infty f_{a}^n(z) a^n(x^\mu) \,, \\
  V_z(x^\mu,z) &= \sum_{n=0}^\infty f_{V_z}^n(z) a^n(x^\mu) \,,
\end{align}
where the 4D modes $a^n(x^\mu)$ satisfy $\Box a^n =m_n^2 a^n$. 
The axion is identified with the massless zero mode. 
The solution for the axion profile was obtained in~\cite{Cox:2019rro}, and for $\lambda=0$ is approximately given by 
\begin{align} \label{eq:axion-profile}
    f_a^0(z) &\simeq \frac{z_{IR}}{\sigma_0} \sqrt{\Delta-1} \left(1 + \frac{g_5^2 k\sigma_0^2}{4\Delta(\Delta-1)}\left( \frac{(\Delta-1)^2}{2\Delta-1} + \frac{z^2}{z_{IR}^2}\left( \left(\frac{z}{z_{IR}}\right)^{2(\Delta-1)} - \Delta\right)\right) \right) \,, \notag \\
    f_{V_z}^0(z) &\simeq \frac{-1}{2\sigma_0\sqrt{\Delta-1}} \frac{z}{z_{IR}}\left( g_5^2 k\sigma_0^2 \left(1-\left(\frac{z}{z_{IR}}\right)^{2(\Delta-1)}\right) \right) \,,
\end{align}
up to corrections of order $(g_5^2k\sigma_0^2/\Delta^2)^2$. Note that the exact profiles are used to obtain our numerical results in later sections. When PQ-violating terms are added on the UV boundary the above profiles are modified in the UV, with the expressions given in \cite{Cox:2019rro}.


\subsection{Bulk Standard Model fermions}

In addition to the bulk $U(1)_{PQ}$ there is also the Standard Model gauge group $SU(3)_c\times SU(2)_L\times U(1)_Y$. The bulk Standard Model gauge bosons have Neumann boundary conditions so that the massless zero modes are identified with the Standard Model gauge bosons (see Ref.~\cite{Gherghetta:2010cj}). Later, we will consider two possibilities for breaking the electroweak gauge symmetry.

The bulk Standard Model gauge group allows for the Standard Model fermions to be located in the bulk. The localization of the zero modes is then responsible for generating the fermion mass hierarchy and will also lead to flavour-dependent axion-fermion couplings. Denoting the 5D $SU(2)_L$ quark doublet field by $Q$ and the singlet fields by $U$, $D$, the bulk fermion action for the quark sector is given by~\cite{Gherghetta:2010cj,Gherghetta:2000qt}
\begin{align}
  S_f = -2\int^{z_{IR}}_{z_{UV}}d^5x\, \sqrt{-g} \Bigg( &\frac{1}{2}\left( \bar{Q}_i \Gamma^M \mathcal{D}_M Q_i - (\mathcal{D}_M \bar{Q}_i) \Gamma^M Q_i \right) + M_{Q_i} \bar{Q}_i Q_i \notag \\
  + &\,\frac{1}{2}\left( \bar{U}_i \Gamma^M \mathcal{D}_M U_i - (\mathcal{D}_M \bar{U}_i) \Gamma^M U_i \right) + M_{U_i} \bar{U}_i U_i \notag \\
  + &\,\frac{1}{2}\left( \bar{D}_i \Gamma^M \mathcal{D}_M D_i - (\mathcal{D}_M \bar{D}_i) \Gamma^M D_i\right) + M_{D_i} \bar{D}_i D_i \Bigg) \,,
\end{align}
where $\Gamma^M = e^M_A \gamma^A = A(z)^{-1}(\gamma^\mu, \gamma^5)$, with $\gamma^5=((\mathbb{1},0),(0,-\mathbb{1}))$, and the fermions carry PQ charges $X_{Q,U,D}$. The 5D masses, $M_X \equiv c_X k$, determine the localization of the chiral zero modes, to be identified with the SM fermions, and $i$ is a flavour index. Decomposing the Dirac spinor $Q_i$ in terms of its Weyl components $Q_i=(Q_{iL}, Q_{iR})^T$, the equation of motion is 
\begin{equation}
    \gamma^\mu\partial_\mu Q_{iL(R)} \mp \partial_z Q_{iR(L)} + \frac{1}{z} \(c_{Q_i} \pm 2\) Q_{iR(L)} = 0 \,.
\end{equation}
To solve this equation, we perform the KK expansion, 
\begin{equation}
    Q_{iL(R)}(x^\mu,z) = \sum_{n=0}^\infty f_{Q_{iL(R)}}^n(z) Q_{iL(R)}^n(x^\mu) \,, 
    \label{KKexpansionBulkFermions}
\end{equation}
where $\slashed\partial Q_{iL(R)}^n = -m_n Q_{iR(L)}^n$, and similarly for $U$ and $D$. After imposing Dirichlet conditions $Q_{iR}=U_{iL}=D_{iL}=0$ on both boundaries, there are chiral zero modes with profiles
\begin{align}
\label{eq:fermionprofiles}
    f_{Q_{iL}}^0(z) &= \mathcal{N}_{Q_i} (k z)^{2-c_{Q_i}} \,, \notag \\
    f_{U_{iR}}^0(z) &= \mathcal{N}_{U_i} (k z)^{2+c_{U_i}} \,, \notag \\
    f_{D_{iR}}^0(z) &= \mathcal{N}_{D_i} (k z)^{2+c_{D_i}} \,.
\end{align}
Normalising the 4D kinetic terms fixes the constants
\begin{equation}
    \mathcal{N}_{X} =\sqrt{\frac{(1\mp 2c_X)k}{2((kz_{IR})^{1\mp 2c_X}-(kz_{UV})^{1\mp 2c_{X}})}}\,,
\end{equation}
where $-(+)$ refers to the left (right) handed profiles.
Similar expressions are obtained in the lepton sector. 


\section{Boundary Higgs fields}
\label{sec:boundaryHiggs}

We first consider a setup with boundary-localized Higgs fields $H_{u,d}$ to construct a 5D model of the DFSZ axion~\cite{Zhitnitsky:1980tq,Dine:1981rt}. The Higgs doublet fields, which transform as $H_{u,d} \sim ({\bf 2}, \mp \frac{1}{2})$ under the $SU(2)_L\times U(1)_Y$ electroweak gauge group, are localized on the UV boundary. They are also charged under the $U(1)_{PQ}$ symmetry with charges $X_{H_u,H_d}$, such that $X_{H_u}+X_{H_d} +2 X_\Phi =0$.  The most general scalar potential on the UV boundary is thus
\begin{align}
\label{eq:UVscalarpotential}
    U_{UV}(\Phi,H_u, H_d) &= \lambda_u (|H_u|^2 - v_u^2)^2 + \lambda_d (|H_d|^2 - v_d^2)^2 + b_{UV} k |\Phi|^2 \nonumber\\
    &+\, (a |H_u|^2 + b |H_d|^2) |\Phi|^2 + c( H_u H_d\Phi^2 + h.c.) \nonumber\\
    &+\, d|H_u H_d|^2 + e |H_u^\dagger H_d|^2\,,
\end{align}
where $H_u H_d = \epsilon_{ij}H_u^i H_d^j$ with $\epsilon_{ij}$ the $SU(2)$ antisymmetric tensor. 

To obtain the axion couplings, we first parametrise the scalar fields by
\begin{equation}
\label{eq:scalarVEVs}
    H_u = \frac{v_u}{\sqrt{2}} 
    e^{i\frac{a_u(x)}{v_u}} 
    \begin{pmatrix} 1\\0\end{pmatrix}, \qquad H_d = \frac{v_{d}}{\sqrt{2}} 
    e^{i\frac{a_d(x)}{v_{d}}} 
    \begin{pmatrix} 0\\1\end{pmatrix}, \qquad \Phi = \eta(z) e^{i a(x,z)} \,,
\end{equation}
where we have ignored the radial components and the electromagnetically-charged NG bosons in $H_{u,d}$. The global 4D $U(1)_{PQ}$ symmetry is a remnant of the 5D local $U(1)_{PQ}$ symmetry and is realised by choosing the 5D gauge transformation parameter $\alpha(x,z) = \alpha_0 f_a^0(z)$, such that the axion zero mode transforms as $a^0(x) \rightarrow a^0(x) + \alpha_0$~\cite{Cox:2019rro}. The 4D PQ current can then be written as
\begin{equation}
    J_\mu^{PQ} = 
        X_\Phi f_a^0(z_{UV})^{-1}{\partial_\mu}a^0 +
        X_{H_u} H_u^\dagger i\overleftrightarrow{\partial_\mu} H_u +
        X_{H_d} H_d^\dagger i\overleftrightarrow{\partial_\mu} H_d + 
        \ldots \,,
\end{equation}
where $H_i^\dagger\overleftrightarrow{\partial_\mu} H_i = \partial_\mu(H_i^\dagger) H_i - H_i^\dagger\partial_\mu H_i$.
The physical 4D axion, $a_4$, is then defined by using the Goldstone theorem $\langle 0| J_\mu^{PQ}|a_4\rangle = i F_a p_\mu$. This gives:
\begin{equation}
\label{eq:axiondefn}
    F_a a_4(x) \equiv X_\Phi f_a^0(z_{UV})^{-1} a^0 + X_{H_u} v_u a_u + X_{H_d} v_d a_d\,,
\end{equation}
where
\begin{equation}
    \sum_i X_i^2 v_i^2 = F_a^2\,,
\end{equation}
with $i=\Phi,H_{u,d}$ and $v_\Phi = f_a^0(z_{UV})^{-1}$. Since $v_{u,d}\ll v_\Phi$ we obtain that $F_a\simeq v_\Phi$.

Similarly, the 4D hypercharge current is given by
\begin{equation}
    J_\mu^Y = 
        Y_u H_u^\dagger i\overleftrightarrow{\partial_\mu} H_u +
        Y_d H_d^\dagger i\overleftrightarrow{\partial_\mu} H_d =
        \frac{1}{2} \partial_\mu(v_u a_u-v_d a_d)\,,
\end{equation}
where $Y_{u,d} = \mp 1/2$ and $a_Z\propto v_u a_u - v_d a_d$ is the NG boson eaten by the $Z$ boson. Requiring orthogonality between the PQ and hypercharge currents, i.e. $\langle 0|J_\mu^Y|a_4\rangle = 0$, leads to the condition
\begin{equation} \label{eq:orthogonality}
    X_{H_u} v_u^2 - X_{H_d} v_d^2 = 0\,.
\end{equation}
Combined with the relation $X_{H_u} + X_{H_d} + 2 X_\Phi= 0$, this fixes the PQ charges of the scalars up to an overall normalization:
\begin{equation}
    X_\Phi=1, \qquad X_{H_u} = -2 \cos^2\beta,\qquad X_{H_d} = -2 \sin^2\beta\,,
\end{equation}
where $\sin\beta=v_u/v$, $\cos\beta=v_d/v$, with the electroweak VEV $v=246$\,GeV. 

In addition to the physical axion defined in \eqref{eq:axiondefn} there is also a heavy axion $A(x)$ that obtains its mass from the $H_u H_d \Phi^2 +h.c.$ term in the boundary potential \eqref{eq:UVscalarpotential}. It is given by
\begin{equation}
    A(x) \propto \frac{a_u(x)}{v_u} + \frac{a_d(x)}{v_d} + 2f_a^0(z_{UV}) a^0(x) +\dots
\end{equation}
The three physical fields $a_4, a_Z$ and $A$ are defined in terms of $a_u,a_d$ and $a^0$. Inverting these relations determines $a_{u,d}$ as a function of the axion field $a_4(x)$. In the limit that $v_\Phi \gg v_{u,d}$ one finds the substitution relations
\begin{equation}
\label{eq:audsub}
    \frac{a_{u,d}}{v_{u,d}} \rightarrow X_{H_u,H_d} \frac{a_4}{F_a}\,.
\end{equation}


\subsection{Axion-fermion couplings with boundary Higgs fields}

To obtain the axion-fermion couplings we first need to specify how the SM fermion masses are generated.
Since the Higgs fields are UV localized, this occurs via Yukawa couplings localized on the UV brane,
\begin{equation}
\label{eq:Yaction}
    S_{\rm Yukawa} = -\int d^4x\, \sqrt{-g_{UV}}\, \frac{1}{k} \bigg( y_{u,ij}^{(5)} \bar{Q}_i U_j H_u + y_{d,ij}^{(5)} \bar{Q}_i D_j H_d + y_{e,ij}^{(5)} \bar{L}_i E_j H_d + \text{h.c.} \bigg) \bigg|_{z_{UV}} \,,
\end{equation}
where $y_{u,d,e}^{(5)}$ are dimensionless 5D Yukawa couplings. The axion couplings to fermions are then obtained by substituting \cref{eq:scalarVEVs} 
into \eqref{eq:Yaction} and using the relations \eqref{eq:audsub}.

\begin{figure}[t]
    \centering
    \includegraphics[width=0.45\textwidth]{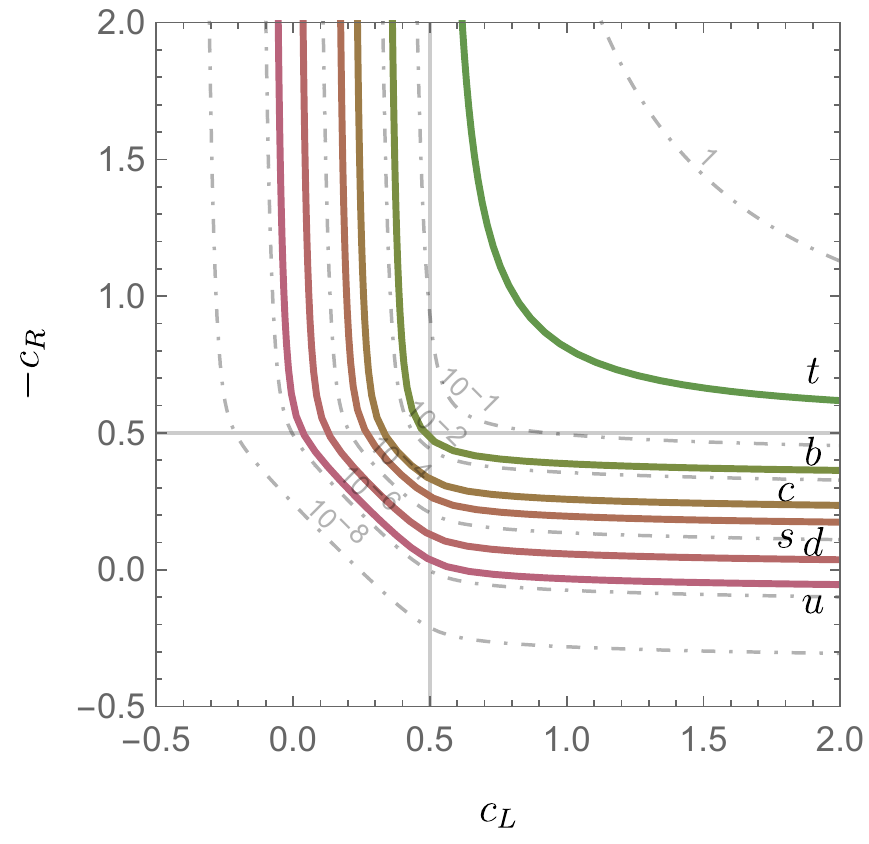} 
    \includegraphics[width=0.45\textwidth]{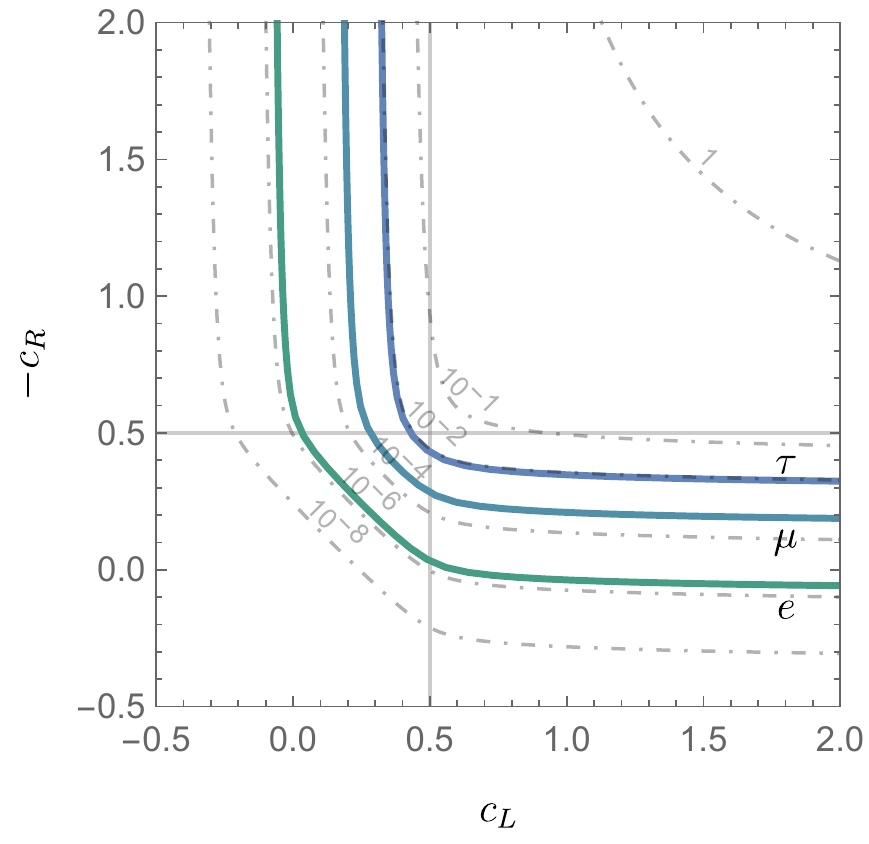} 
    \caption{Contours of the quark (left) and charged lepton (right) masses for the boundary Higgs case as a function of $c_L$ and $-c_R$  with $\tan \beta = 3$, and $k z_{\rm IR} = 10^{10}$. The $y_{u,d,\ell}^{(5)}$ are randomly generated diagonal 5D Yukawa matrices drawn from a log-normal distribution with $\mu = 0$ and $\sigma = 0.3$. The contours correspond to the median of the generated set $(c_L, -c_R)$.}
    \label{fig:masses-boundary}
\end{figure}

By performing the following field redefinitions on the fermion zero modes,
\begin{equation}
\label{eq:fermiontrans}
    u_i\rightarrow e^{i\gamma_5 X_{H_u} \frac{a_4}{2F_a}} u_i, \qquad d_i\rightarrow e^{i\gamma_5 X_{H_d} \frac{a_4}{2F_a}} d_i,\qquad e_i\rightarrow e^{i\gamma_5 X_{H_d} \frac{a_4}{2F_a}} e_i\,,
\end{equation}
the axion field can be removed from the mass terms to give the 4D effective Lagrangian
\begin{equation}
\label{eq:massLag}
    {\cal L}_{4D} \supset -m_u^{ij} {\bar u}_{iL} u_{jR} -m_d^{ij} {\bar d}_{iL} d_{jR} -m_e^{ij} {\bar e}_{iL} e_{jR} + \text{h.c.} \,,
\end{equation}
where 
\begin{equation}
\label{eq:boundarymassrelation}
    m_u^{ij} = y_{u,ij}^{(5)} \frac{v_u}{\sqrt{2}k} f_{Q_{iL}}^0(z_{UV}) f_{U_{jR}}^0(z_{UV}) \,,
\end{equation}
and similarly for $m_{d,e}^{ij}$. The mass matrix $m_u^{ij}$ can be diagonalized by the singular value decomposition $A_L^{u\dagger} m_u^{ij} A_R^u = m_{u_i}$, where $A_{L,R}^u$ are unitary matrices and $m_{u_i}$ is the diagonal mass matrix containing the up-type masses. For a $3\times 3$ matrix $y^{(5)}_u$ with anarchic elements of order one, a 4D Yukawa coupling hierarchy is generated from the overlap of the bulk profiles~\cite{Gherghetta:2000qt}. The bulk mass parameters, $c_i$, are then constrained by the quark and charged lepton masses, as shown in \cref{fig:masses-boundary}. Further details are given in \cref{app:SMflavour}. 

The fermion kinetic terms are not invariant under the redefinitions \eqref{eq:fermiontrans} and generate derivative couplings of the axion to fermions. For the up-type quarks, these are given by 
\begin{align}
    &i\int d^4x\, \frac{X_{H_u}}{2F_a}\partial_\mu a_4\, \left(  -{\bar u}_{iL} (A_{L}^u A_L^{u\dagger})_{ij}\gamma^\mu u_{jL} + {\bar u}_{iR} (A_{R}^u A_R^{u\dagger})_{ij}\gamma^\mu u_{jR} \right)\,,\nonumber\\
    &\equiv i\int d^4 x~\left(\frac{\partial_\mu a_4}{2 F_a}\, c^A_u {\bar u_i} \gamma^\mu  \gamma_5 \, u_i \right)\,,
\end{align}
where $c^A_u = - X_{H_u}$. Similarly for the down-type quarks and leptons. Thus, for boundary-localized Higgs fields, the vector couplings vanish and the axial-vector couplings are flavour-diagonal. The current experimental constraints on the $c^{A}$ are given in Ref.~\cite{MartinCamalich:2020dfe}. The redefinitions \eqref{eq:fermiontrans} also induce axion couplings to gluons and photons, as in the standard DFSZ model. A more detailed investigation of the axion-gluon/photon couplings in the bulk Higgs case will be discussed below.


\section{Bulk Higgs fields}\label{bulkHiggsSection}

Next, we consider the case of bulk Higgs fields, and show that this leads to flavour off-diagonal axion-fermion couplings. The Higgs fields $H_u, H_d$ still have a UV boundary potential given by \eqref{eq:UVscalarpotential} but now they propagate in the bulk. The bulk action is 
\begin{align} 
\label{eq:5D_Higgsaction}
    S_H &= 2\int^{z_{IR}}_{z_{UV}} d^5x\, 
    \sqrt{-g} \left( -\big(\mathcal{D}^M H_{u,d}\big)^\dagger
    \big(\mathcal{D}_M H_{u,d}\big) - m_{H_{u,d}}^2 H_{u,d}^\dagger H_{u,d} \right. \notag \\
    &- \frac{1}{2g_Y^2\xi_Y} \left( g^{\mu\nu}\partial_\mu B_\nu - \xi_Y g_Y^2 \left( Y_{H_u} v_u a_u + Y_{H_d} v_d a_d \right) \right)^2 \Big) \notag \\
    &- \int d^4x\, \sqrt{-g_4} \, U_{UV}(\Phi, H_u,H_d) \,,
\end{align}
where $\mathcal{D}_M= \partial_M - i X_{H_{u,d}} V_M + \dots$, and $\xi_Y$ is a $U(1)_Y$ gauge-fixing parameter. Note that $a_{u,d}$ will also contribute to the $U(1)_{PQ}$ gauge fixing term in eq.~\eqref{eq:5D_action}.
The scalar fields can be parametrised as
\begin{equation}
\label{eq:bulkscalarVEVs}
    H_u = \frac{v_u}{\sqrt{2}} 
    e^{i \frac{a_u(x,z)}{v_u}} 
    \begin{pmatrix} 1\\0\end{pmatrix}, \qquad H_d = \frac{v_d}{\sqrt{2}} 
    e^{i \frac{a_d(x,z)}{v_d}} 
    \begin{pmatrix} 0\\1\end{pmatrix}, \qquad \Phi = \eta(z) e^{i a(x,z)} \,,
\end{equation}
where $a_{u,d}(x,z)$ are the neutral NG bosons propagating in the bulk, and the radial components and the electromagnetically-charged NG bosons in $H_{u,d}$ have again been ignored. In general, the bulk VEVs can have nontrivial $z$-dependence (i.e. $v_{u,d}=v_{u,d}(z)$), but for simplicity we assume they are constant. This requires adding appropriate bulk and IR boundary mass terms for $H_{u,d}$, similar to the bulk Higgs setup considered in Ref.~\cite{Cacciapaglia:2006mz}. The 4D electroweak VEV is then approximately $v^2 \approx (v_u^2+v_d^2)/k$, assuming $z_{UV}=1/k$ (note that the 5D fields in \eqref{eq:bulkscalarVEVs} have canonical mass dimension $3/2$).

It is convenient to define the new fields
\begin{align}
    a_Y &= \frac{1}{N_Y} \left( Y_{H_u} v_u a_u + Y_{H_d} v_d a_d \right) \,, \\
    a_X &= \frac{1}{N_X} \left( X_{H_u} v_u a_u + X_{H_d} v_d a_d \right) \,,
\end{align}
where $N_Y=\sqrt{Y_{H_u}^2 v_u^2 + Y_{H_d}^2 v_d^2}$ and $N_X=\sqrt{X_{H_u}^2 v_u^2 + X_{H_d}^2 v_d^2}$. The PQ charges can be chosen such that these two combinations are orthogonal:
\begin{equation}
    Y_{H_u} X_{H_u} v_u^2 + Y_{H_d} X_{H_d} v_d^2 = 0 \,.
\end{equation}
Combining this with the condition $X_{H_u} + X_{H_d} + 2X_\Phi = 0$ (with $X_\Phi=1$ and $Y_{H_{u,d}} = \mp 1/2$) yields the relations
\begin{equation}
    X_{H_u} = \frac{-2 v_d^2}{v_u^2 + v_d^2} \,, \qquad X_{H_d} = \frac{-2 v_u^2}{v_u^2 + v_d^2} \,,
\end{equation}
and hence $N_X=2 v_u v_d / \sqrt{v_u^2 + v_d^2}$. The equations of motion for $a_Y$ and $a_X$ then decouple:
\begin{equation} \label{eq:a_Y-eom}
  A^3\,\Box a_Y + \partial_z\left(A^3 \partial_z a_Y \right) - \xi_Y A^5 g_Y^2 N_Y^2 a_Y = 0 \,,
\end{equation}
\begin{equation} \label{eq:a_X-eom}
   A^3\,\Box a_X + \partial_z\left(A^3 \left( \partial_z a_X - N_X V_z \right)\right) + \xi_{PQ} A^5 N_X \left(A^{-3} \partial_z\left(A V_z\right) -g_5^2\left( X_\Phi \eta^2 a + N_X a_X \right) \right) = 0 \,. 
\end{equation}
Note that in deriving these equations we have used the fact that $v_{u,d}$ are $z$-independent. The boundary conditions are
\begin{align}
    \left( \pm 2 A^3 \partial_z a_Y - A^4\frac{\delta U}{\delta a_Y} \right)\bigg|_{z_{UV},z_{IR}} &= 0 \,, \\
    \label{eq:axbc}
    \left( \pm 2 A^3 \left(\partial_z a_X - N_X V_z\right) -A^4\frac{\delta U}{\delta a_X} \right)\bigg|_{z_{UV},z_{IR}} &= 0 \,,
\end{align}
with the relevant part of the UV boundary potential given by
\begin{equation}
    U_{UV} \supset c v_u v_d \eta^2 \cos\bigg(2a - \frac{\sqrt{v_u^2 + v_d^2}}{v_u v_d} a_X \bigg)\bigg|_{z_{UV}} \,.
\end{equation}

It is convenient to work in unitary gauge for $U(1)_Y$ ($\xi_Y\to\infty$) since then $a_Y\to0$. For $a_X$, we perform the KK expansion, 
\begin{equation}\label{aXKKexpansion}
 a_X(x^\mu,z) = \sum_{n=0}^\infty f_{a_X}^n(z) a^n(x^\mu) \,.
\end{equation}
Note that $a^n(x^\mu)$ are the same 4D modes as in eq.~\eqref{eq:KK-expansion}. To solve eq.~\eqref{eq:a_X-eom} we expand to first-order in $v_{u,d}/k^{3/2}$ (higher-order terms are negligibly small). This allows us to neglect terms proportional to $a_X$ in the equation of motion for $a$ and $V_z$. Focusing on the massless mode, we can then continue to use the solutions for $f_a^0$ and $f_{V_z}^0$ in \eqref{eq:axion-profile}. These massless profiles satisfy  $A^{-3} \partial_z(A f_{V_z}^0) = g_5^2 X_\Phi \eta^2 f_a^0$. Using this relation, \cref{eq:a_X-eom} reduces to
\begin{equation}
    \partial_z\left(A^3 \left( \partial_z f_{a_X}^0 - N_X f_{V_z}^0 \right)\right) = 0 \,,
\end{equation}
for the massless mode. Imposing the IR boundary condition \eqref{eq:axbc} enforces $\partial_z f_{a_X}^0(z) = N_X f_{V_z}^0(z)$. The UV boundary condition then becomes
\begin{equation}
    - c A^4 \eta^2 \sqrt{v_u^2 + v_d^2} \left( 2f_a^0 - \frac{\sqrt{v_u^2 + v_d^2}}{v_u v_d} f_{a_X}^0 \right) \Bigg|_{z_{UV}} = 0 \,.
\end{equation}
The final solution is
\begin{equation} \label{eq:a_X-solution}
    f_{a_X}^0(z) = \frac{2 v_u v_d}{\sqrt{v_u^2 + v_d^2}} \left( f_a^0(z_{UV}) + \int_{z_{UV}}^{z} dz' \, f_{V_z}^0(z') + \mathcal{O}(v_{u,d}^2/k^3) \right) \,.  
\end{equation}
This solution is strictly valid only for an exactly massless zero mode. However, it is expected to approximately hold even in the presence of explicit PQ breaking on the UV boundary with the replacement $f_a^0(z_{UV}) \to f_{a,\cancel{PQ}}^0(z_{UV})$, where the latter quantity is the boundary value of the exact massive profile.\footnote{We have confirmed that with this replacement \eqref{eq:a_X-solution} holds exactly when $g_5=0$ and $z_{IR} \gg z_{UV}$.} The reason is that, in the limit $z_{IR} \gg z_{UV}$, the exact massive profiles $f_{a,\cancel{PQ}}^0$ and $f_{V_z,\cancel{PQ}}^0$ closely match the massless solutions everywhere except very close to the UV brane, where $f_{a,\cancel{PQ}}^0$ becomes highly suppressed~\cite{Cox:2019rro}.
It is therefore convenient to write
\begin{equation} \label{eq:a_X-profile}
    f_{a_X}^0(z) = \frac{2 v_u v_d}{\sqrt{v_u^2 + v_d^2}} \left( \widehat{f}_{a_X}^0(z) + \mathcal{O}(v_{u,d}^2/k^3) \right) \,,
\end{equation}
with
\begin{equation} \label{eq:fhat-def}
    \widehat{f}_{a_X}^0(z) = 
    \begin{cases} 
      f_a^0(z) \,, & m_0=0 \,, \\
      f_a^0(z) - f_a^0(z_{UV}) + f_{a,\cancel{PQ}}^0(z_{UV}) \,, & m_0 \neq 0 \,,
   \end{cases}
\end{equation}
where the equality should be understood as approximate when $m_0 \neq 0$. In going from \eqref{eq:a_X-solution} to \eqref{eq:fhat-def} we have used that $f_{V_z}^0=\partial_z f_a^0$. From now on, we also approximate $f_{a,\cancel{PQ}}^0(z_{UV}) \approx 0$.

Finally, transforming back to $a_{u,d}$ we obtain
\begin{equation} \label{eq:bulksub}
    \frac{a_{u,d}}{v_{u,d}} = X_{H_{u,d}} \widehat{f}_{a_X}^0(z)\, a^0(x^\mu) + \ldots \,,
\end{equation}
where `$\ldots$' contains the heavier modes.

\subsection{Axion-fermion couplings with bulk Higgs fields}
\label{sec:axionfermionBulk}

\begin{figure}[ht]
    \centering
    \includegraphics[width=0.47\textwidth]{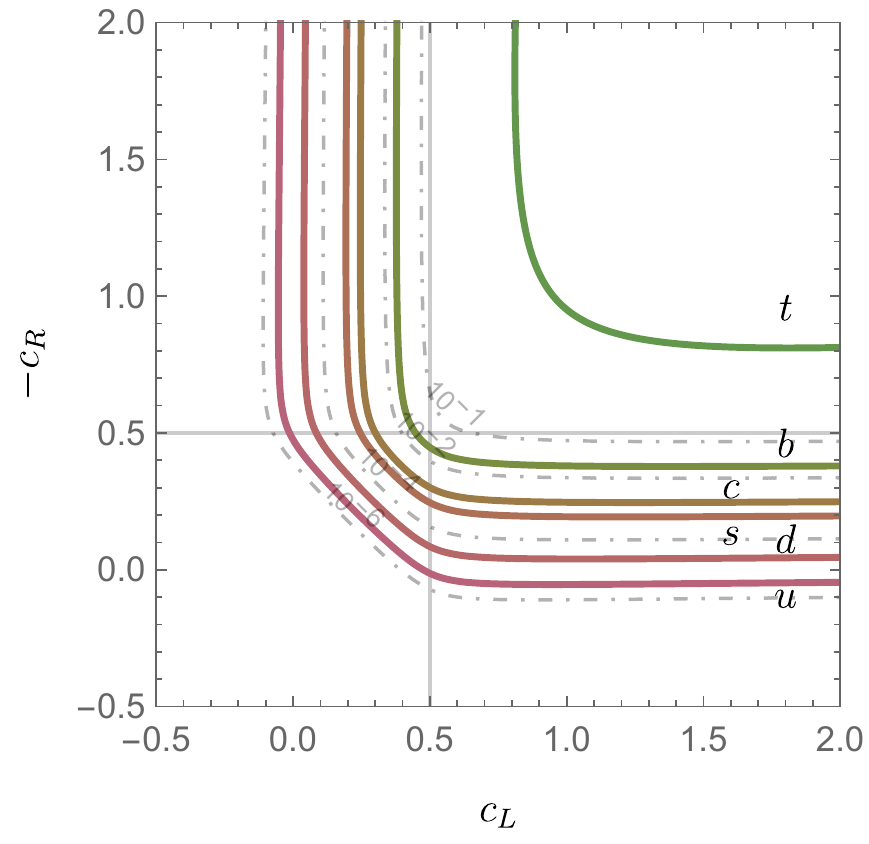} 
    \includegraphics[width=0.47\textwidth]{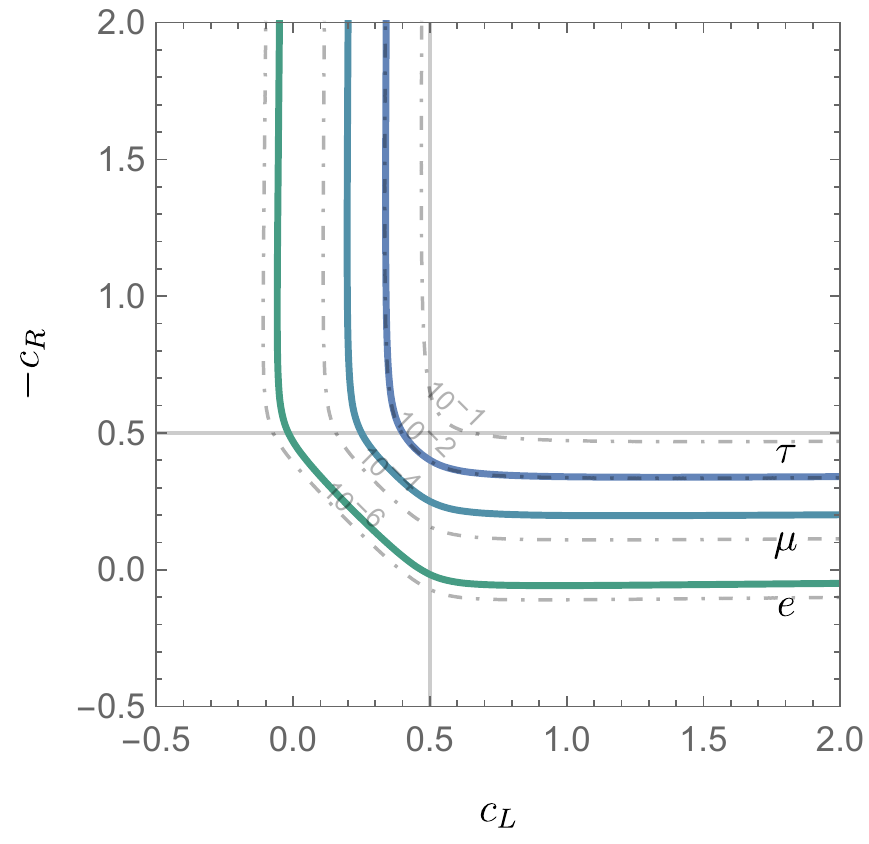} 
    \caption{Contours of the quark (left) and charged lepton (right) masses for the bulk Higgs case using Eq.~\eqref{eq:5Dmassrelation}, as a function of $c_L$ and $-c_R$  with $\tan \beta = 3$, and $k z_{\rm IR} = 10^{10}$. The $y_{u,d,\ell}^{(5)}$ are randomly generated diagonal 5D Yukawa couplings drawn from a log-normal distribution with $\mu = 0$ and  $\sigma = 0.3$. The contours indicate the medians of the generated set $(c_L, -c_R)$}
    \label{fig:masses-bulk}
\end{figure}

To obtain the axion-fermion couplings we consider the bulk Yukawa interactions
\begin{equation} \label{eq:bulkYaction}
  S_{\rm Yukawa} = -2\int^{z_{IR}}_{z_{UV}} d^5x\, \sqrt{-g} \, \frac{1}{\sqrt{k}} \bigg( y_{u,ij}^{(5)} \bar{Q}_i U_j H_u + y_{d,ij}^{(5)} \bar{Q}_i D_j H_d + y_{e,ij}^{(5)} \bar{L}_i E_j H_d + \text{h.c.} \bigg) \,,
\end{equation}
where $y_{u,d,e}^{(5)}$ are dimensionless 5D Yukawa couplings. Focusing for now on the quark sector, and substituting \cref{eq:bulkscalarVEVs} into \eqref{eq:bulkYaction} gives
\begin{equation} \label{eq:bulkYaction2}
   -2\int^{z_{IR}}_{z_{UV}} d^5x\, \sqrt{-g} \, \frac{1}{\sqrt{k}} \left( y_{u,ij}^{(5)} \frac{v_u}{\sqrt{2}} {\bar Q}_{u_i} U_j e^{i \frac{a_u(x,z)}{v_u}} + y_{d,ij}^{(5)} \frac{v_d}{\sqrt{2}} {\bar Q}_{d_i} D_j e^{i \frac{a_d(x,z)}{v_d}} + \text{h.c.} \right)\,,
\end{equation}
where $Q=(Q_u, Q_d)$ denote the components of the $SU(2)_L$ quark doublet.
We proceed by considering just the up-type quarks. Similar expressions follow for the down-type quarks and leptons. The fermion zero-mode mass matrix for the up-type quarks is
\begin{equation} \label{eq:5Dmassrelation}
 m_u^{ij} = y_{u,ij}^{(5)} \frac{\sqrt{2}v_u}{\sqrt{k}} \int_{z_{UV}}^{z_{IR}} \frac{dz}{(kz)^5}~f_{Q_{iL}}^0(z) f_{U_{jR}}^0(z) \,,
\end{equation}
which is again diagonalised by $A_L^u m_u^{ij} A_R^{u\dagger} = m_{u_i}$. As in the boundary Higgs case,  the bulk mass parameters, $c_i$, are constrained by the quark and charged lepton masses, as shown in \cref{fig:masses-bulk}. Further details are provided in \cref{app:SMflavour}. 

The $a_u$ dependence in \cref{eq:bulkYaction2} can be removed via a 5D field redefinition of the form
\begin{equation} \label{eq:5Dfermionredef}
    Q_{u_i}(x,z) \rightarrow e^{i \beta \frac{a_u(x,z)}{v_u}} Q_{u_i}(x,z), \quad U_i(x,z) \rightarrow e^{i (\beta - 1) \frac{a_u(x,z)}{v_u}} U_i(x,z)\,,
\end{equation}
where $\beta$ is an arbitrary parameter. The 5D kinetic terms are not invariant under this transformation, giving rise to the terms
\begin{multline} \label{eq:5D-aff-coupling}
    -2i\int_{z_{UV}}^{z_{IR}} d^5x \sqrt{-g}\, \left(\partial_M \frac{a_u}{v_u}\right) \bigg( \frac{1}{2} \left( \bar{Q}_{u_i} \Gamma^M Q_{u_i} - \bar{U}_i \Gamma^M U_i \right) \\
    + \left(\beta - \frac{1}{2}\right) \left( \bar{Q}_{u_i} \Gamma^M Q_{u_i} + \bar{U}_i \Gamma^M U_i \right) \bigg) \,.
\end{multline}
Restricting to the zero-modes, these terms give the axion-fermion couplings. However, care must be taken with the term on the second line, which depends on the choice of field redefinition. Notice that after integrating by parts (the boundary term vanishes) this term takes the form
\begin{equation}
    i(2\beta -1) \int_{z_{UV}}^{z_{IR}} d^5x\, \frac{a_u}{v_u} \partial_M J_V^M \,, \qquad J_V^M = \sqrt{-g} \left( \bar{Q}_{u_i} \Gamma^M Q_{u_i} + \bar{U}_i \Gamma^M U_i \right) \,.
\end{equation}
Since $J_V^M$ is a classically conserved current, any effects from this term must be proportional to the (boundary-localized) weak anomaly. The $\beta$-dependence is then cancelled by the transformation of the path-integral measure under \eqref{eq:5Dfermionredef}.

Returning to the terms in the first line of \cref{eq:5D-aff-coupling} and restricting to the zero-modes gives
\begin{equation}
    -iX_{H_u} \int_{z_{UV}}^{z_{IR}} d^5x\, A^4 (\partial_\mu a^0) \widehat{f}_{a_X}^0 \left( \bar{u}_{iL} (f^0_{Q_{iL}})^2 \gamma^\mu u_{iL} - \bar{u}_{iR} (f^0_{U_{iR}})^2 \gamma^\mu u_{iR} \right) \,,
\end{equation}
where we have used \eqref{eq:bulksub} and $a^0(x)$ is identified with the axion to ${\cal O}(v/F_a)$. Integrating over the profiles and rotating to the fermion mass basis we obtain the 4D effective action
\begin{equation} \label{eq:cAcV-bulk}
    S_{4D} \supset i\int d^4x\, \frac{\partial_\mu a^0}{2F_a} \( \bar{u}_i \gamma^\mu \left( (c_u^V)_{ij} - (c_u^A)_{ij} \gamma^5 \right) u_j \) \,,
\end{equation}
where
\begin{equation} \label{eq:cVA_bulk}
\frac{1}{(F_u^{V,A})_{ij}}\equiv \frac{(c_u^{V,A})_{ij}}{F_a} 
    = X_{H_u} \int_{z_{UV}}^{z_{IR}} \frac{dz}{(kz)^4}\, \widehat{f}_{a_X}^0 \left( (A^u_R)_{ik} (f^0_{U_{kR}})^2 ({A^u_R}^\dagger)_{kj} \mp (A^u_L)_{ik} (f^0_{Q_{kL}})^2 ({A^u_L}^\dagger)_{kj} \right) \,.
\end{equation}
Repeating the above steps leads to analogous expressions for the down-type quarks and charged leptons. Note that, following a similar argument to above, the flavour-diagonal vector couplings $(c^V)_{ii}$ are unphysical up to weak anomalies. Furthermore, the on-shell axion couplings are proportional to $(c^{V,A})_{ij} (m_i\mp m_j)$. 

We see that with the Higgs located in the bulk both the vector, $c^V$,  and axial-vector, $c^A$, couplings are non-zero and are flavour off-diagonal. These couplings depend on the mixing matrices, $A_{L,R}$, and the 5D bulk mass parameters $c_i$ in \eqref{eq:fermionprofiles}, which are constrained by a fit to the Standard Model fermion masses and CKM/PMNS matrices, as detailed in \cref{app:SMflavour}. In the following two subsections, we discuss the behaviour of the couplings \eqref{eq:cVA_bulk} both for the massless axion and in the presence of explicit PQ breaking on the UV boundary.


\subsubsection{Massless axion}

\begin{figure}[ht]
    \centering
    \includegraphics[width=0.55\textwidth]{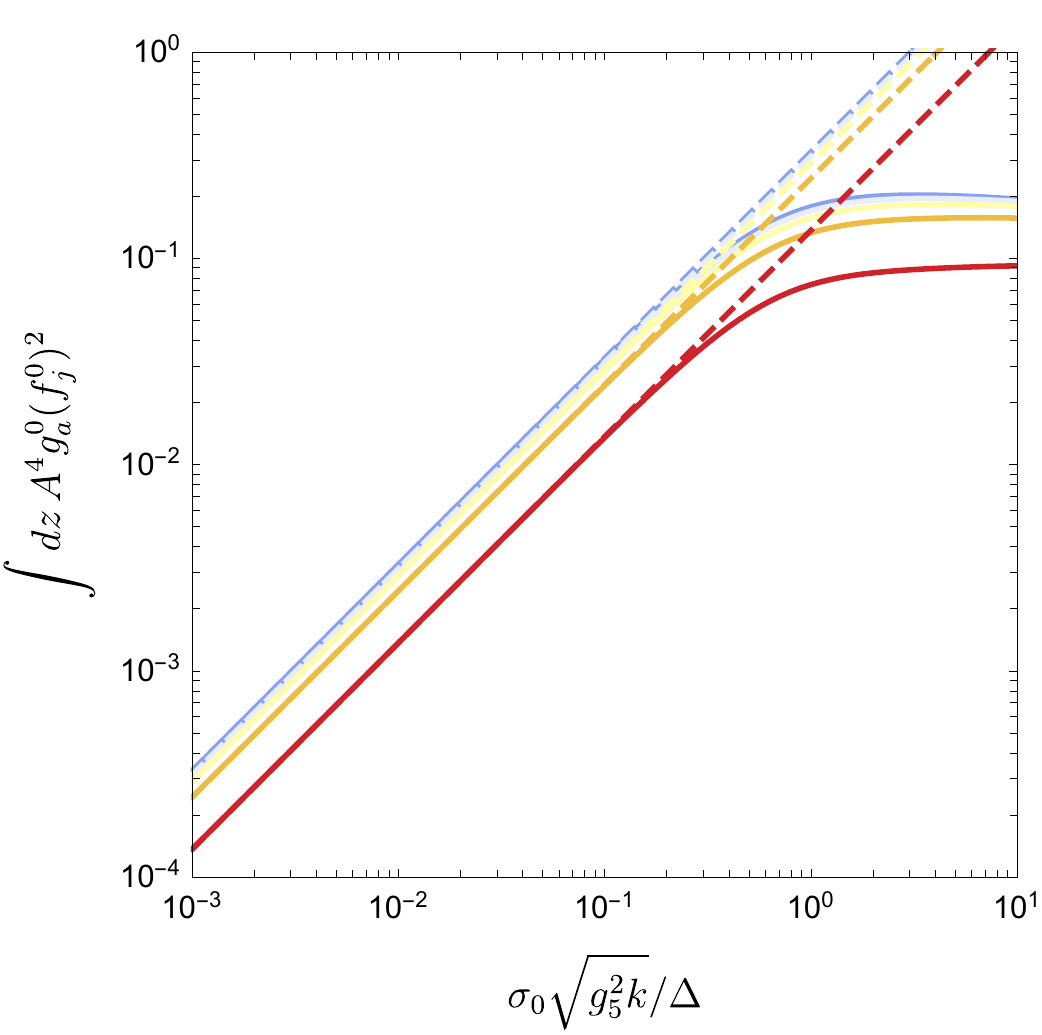} 
    \caption{The overlap integral between the $z$-dependent part of the axion profile $g_a^0(z)$ and the left/right-handed quark profiles $(f_j^0)^2$ as a function of $\sigma_0 \sqrt{g_5^2k} / \Delta$, with $c_L (-c_R)$ ranging from $-3$ (upper) to $0$ (lower) in steps of $1$. The solid lines use the exact massless axion profile, while the dashed lines correspond to \cref{eq:overlapint}. We have fixed $\Delta = 10$, $g_5^2k = 1$ and the value for $k z_{IR}$ is adjusted to keep $F_a = 10^9$\,GeV.}
    \label{fig:overlap-integral}
\end{figure}

In the absence of explicit PQ breaking on the UV brane, the axion is massless (up to QCD effects) and the profiles are given by \cref{eq:axion-profile}. In this case we have that $\widehat{f}_{a_X}^0(z)=f_a^0(z)$. The $f_a^0(z)$ profile is approximately constant, up to corrections of order $g_5^2k\sigma_0^2/\Delta^2$, and it is convenient to parametrize it as
\begin{equation} \label{eq:f_a-zpart}
    f_a^0(z) = \frac{1}{F_a} \left( 1 + g_a^0 (z) \right) \,,
\end{equation}
where $g_a^0(z)$ contains the $z$-dependence, and we have identified the decay constant $F_a=f_a(z_{UV})^{-1}\simeq\sigma_0/(z_{IR}\sqrt{\Delta-1})$. Substituting this into \cref{eq:cVA_bulk} and using the fact that $A_{L,R}$ are unitary matrices leads to
\begin{align} 
    (c^{A}_u)_{ij} &= X_{H_u}  
    \bigg( \frac{1}{2} \delta_{ij} + \int_{z_{UV}}^{z_{IR}} \frac{dz}{(kz)^4}\, g_a^0(z) \left( (A_{R}^u)_{ik}  (f_{U_{kR}}^0)^2 (A_R^{u\dagger})_{kj} + (A_{L}^u)_{ik}  (f_{Q_{kL}}^0)^2 (A_L^{u\dagger})_{kj} \right) \bigg) \,, \notag \\
(c^{V}_u)_{ij} &= X_{H_u}  
     \int_{z_{UV}}^{z_{IR}} \frac{dz}{(kz)^4}\, g_a^0(z) \left( (A_{R}^u)_{ik}  (f_{U_{kR}}^0)^2 (A_R^{u\dagger})_{kj} - (A_{L}^u)_{ik}  (f_{Q_{kL}}^0)^2 (A_L^{u\dagger})_{kj} \right) \,.
\end{align}
The first term in the expression for $c^{A}_u$ gives the leading contribution to the diagonal couplings which are therefore similar to the boundary Higgs case. The off-diagonal couplings, on the other hand, involve overlap integrals of $g_a^0(z)$ with the fermion profiles. These integrals take a particularly simple form if the fermion profile satisfies $c_L<1/2$ or $-c_R<1/2$. Using the approximate axion profile in \cref{eq:axion-profile}, we then obtain
\begin{equation} \label{eq:overlapint}
    \int \frac{dz}{(kz)^4}\, g_a^0\, (f_{j}^0)^2 = 
    - \frac{\Delta(2c_j - 1)(2c_j - 2\Delta - 3)}{8(2c_j - 3)(2c_j - 2\Delta - 1)} \frac{g_5^2 k \sigma_0^2}{\Delta^2} + \mathcal{O}\(\(\frac{g_5^2k\sigma_0^2}{\Delta^2}\)^2\)\,.
\end{equation}
Conversely, if $c_L>1/2$ or $-c_R>1/2$ the integral is suppressed by powers of $kz_{IR}$. From figure~\ref{fig:masses-bulk}, we see that for all fermions except the top quark, \cref{eq:overlapint} is always valid for either the left- or right-handed profile. The overlap integral is plotted in figure~\ref{fig:overlap-integral} using the exact massless axion profile (the dashed lines correspond to \cref{eq:overlapint}). Notice that even when $g_5\sqrt{k} \sigma_0/\Delta\sim\mathcal{O}(1)$ the value of the overlap integral is $\sim0.1$, which results in a suppression of the off-diagonal couplings relative to $F_a$. The off-diagonal couplings are further suppressed by the off-diagonal elements of the mixing matrices.

\subsubsection{Massive axion}

In the presence of explicit PQ breaking on the UV brane $\widehat{f}_{a_X}^0(z) \approx f_a^0(z) - f_a^0(z_{UV}) = g_a^0(z)$. We then obtain the couplings
\begin{equation}
        (c^{V,A}_u)_{ij} = X_{H_u} \int_{z_{UV}}^{z_{IR}} \frac{dz}{(kz)^4}\, g_a^0(z) \left( (A_{R}^u)_{ik}  (f_{U_{kR}}^0)^2 (A_R^{u\dagger})_{kj} \mp (A_{L}^u)_{ik}  (f_{Q_{kL}}^0)^2 (A_L^{u\dagger})_{kj} \right) \,. 
        \label{eq:massiveoverlapint}
\end{equation}
Here, the diagonal couplings $(c^A_u)_{ii}$ are also suppressed, relative to $F_a$, by overlap integrals of $g_a^0(z)$ with the fermion profiles. The off-diagonal couplings are still expected to be smaller, due to the suppression from the off-diagonal elements of the mixing matrices. Hence, the off-diagonal couplings are approximately the same in the massive and massless cases, whereas the diagonal couplings are different. 


\subsubsection{Numerical results}

The axion-fermion couplings are obtained by numerically evaluating the integral expression \eqref{eq:massiveoverlapint}, as well as the corresponding expressions for the down-type quarks and charged leptons. The procedure for determining the mixing matrices $A_{L,R}^{u,d,\ell}$ is given in \cref{app:quark-mixing}, with the bulk mass parameters constrained by a fit to the quark and charged lepton masses (see figure~\ref{fig:masses-bulk}) and the CKM/PMNS mixing matrices. We also assume $z_{\rm UV}=1/k$ and $k\sim M_P$ for all plots.

\begin{figure}[t]
    \centering
    \includegraphics[width=0.45\textwidth]{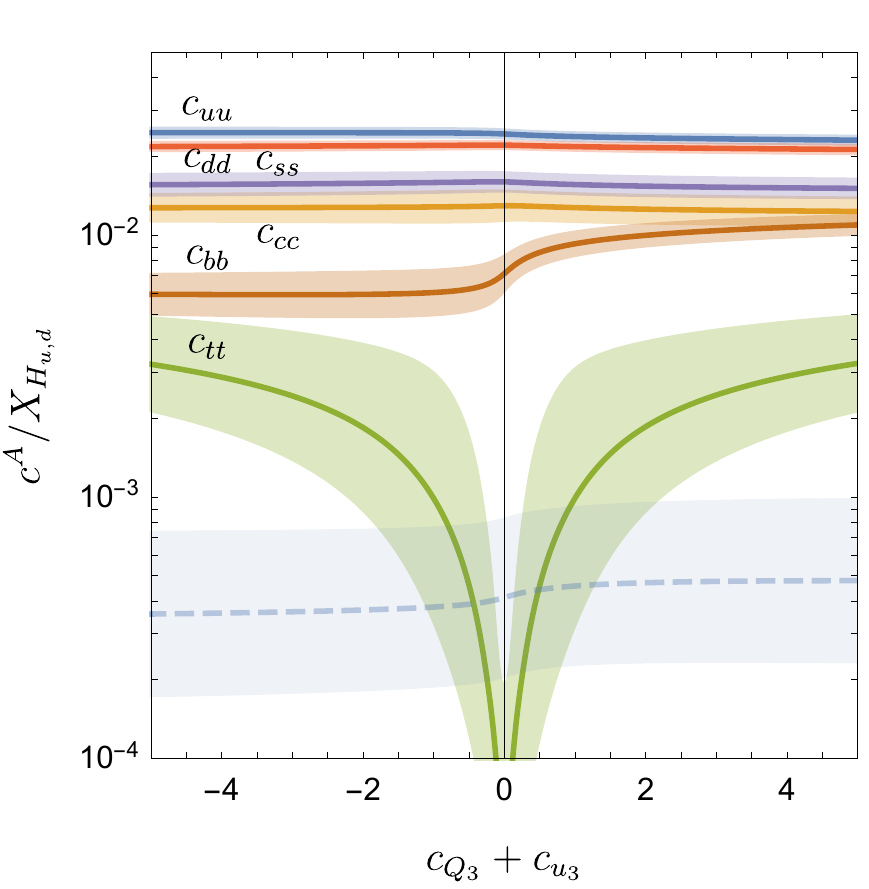} 
    \includegraphics[width=0.45\textwidth]{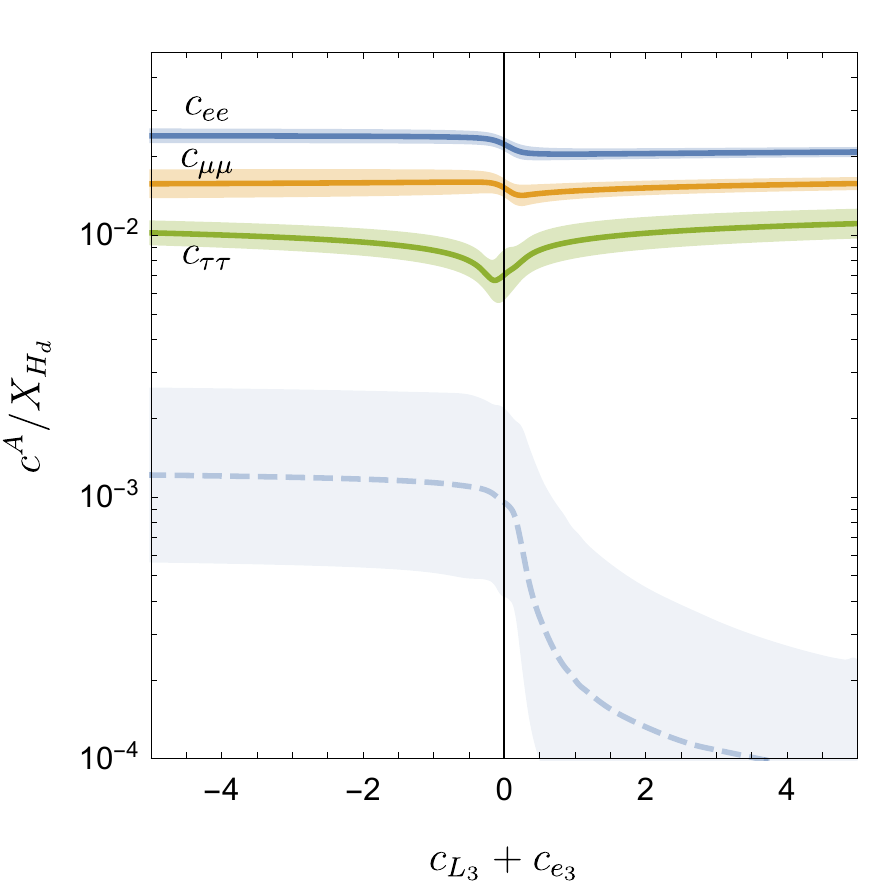} 
    \caption{Left: absolute values of the diagonal axion-quark couplings $c^A_{uu}$ (blue), $c^A_{cc}$ (orange) and $c^A_{tt}$ (green) in units of $X_{H_u}$, and $c^A_{dd}$ (red), $c^A_{ss}$ (purple) and $c^A_{bb}$ (tan) in units of $X_{H_d}$ as functions of $c_{Q_3} + c_{u_3}$. Right: absolute values of the diagonal axion-charged lepton couplings  $c^A_{ee}$ (blue), $c^A_{\mu \mu}$ (orange) and $c^A_{\tau \tau}$ (green) in units of $X_{H_d}$ as functions of $c_{L_3} + c_{e_3}$. We fix $kz_{\rm IR} = 10^{10}, g^2_5 k = 1, \Delta = 10$ and $\sigma_0 = 3$, corresponding to $F_a\simeq 10^9$\,GeV. The curves and bands depict the mean and standard deviation of $\log_{10} F^V$ obtained from a scan over anarchic 5D Yukawa couplings. The dashed line shows (left) $c^A_{uc}$  and (right) $c^A_{e\mu}$ for reference. }
    \label{fig:cv-diag}
\end{figure}

\begin{figure}[t]
    \centering
    \includegraphics[width=0.47\textwidth]{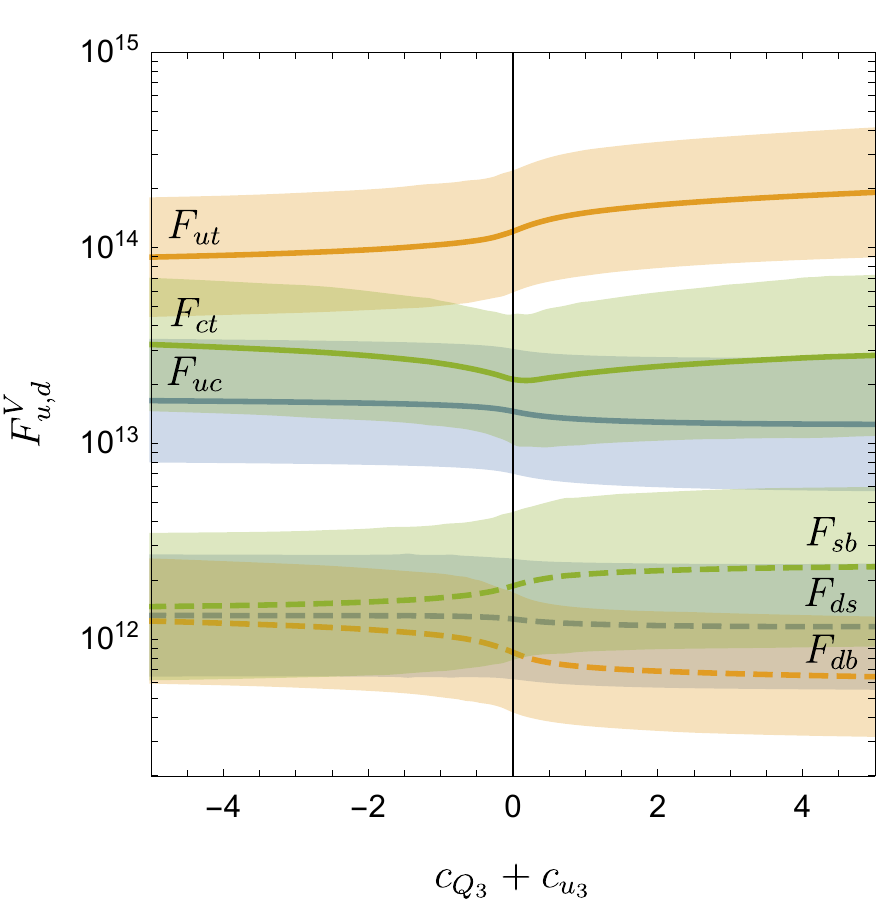} 
    \includegraphics[width=0.47\textwidth]{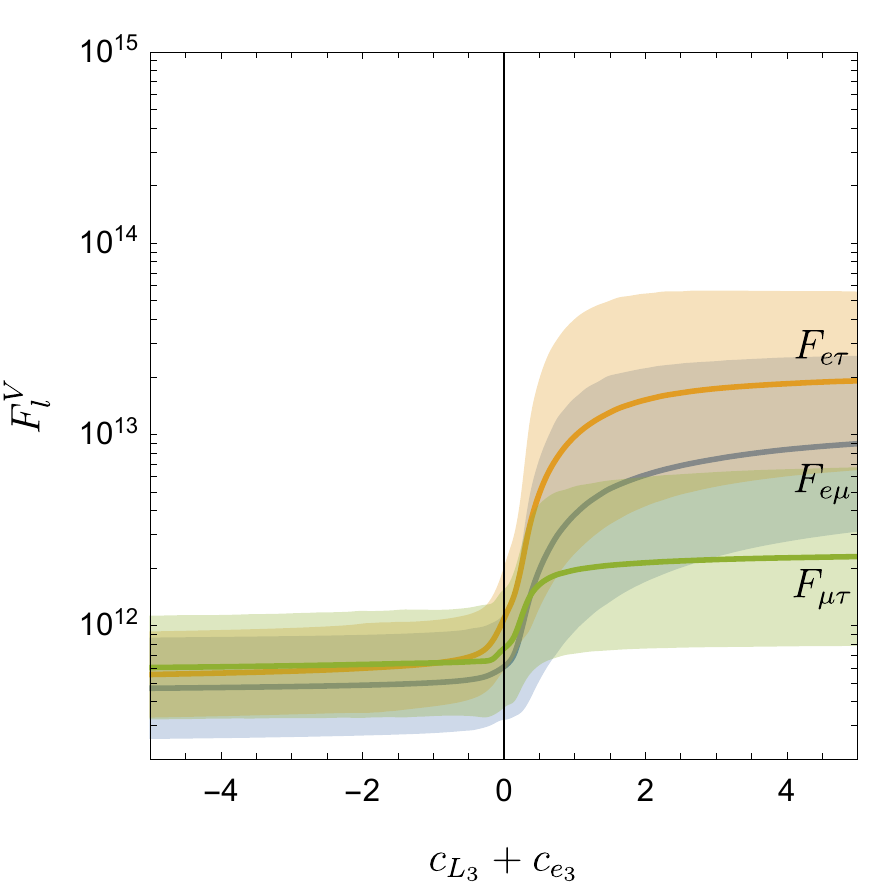} 
    \caption{Absolute values of the off-diagonal elements of the axion-quark coupling matrix $F^V_{u}$ (left, solid) and $F^V_{d}$ (left, dashed) as functions of $c_{Q_3} + c_{u_3}$, and $F^V_{l}$ (right) as a function of $c_{L_3} + c_{e_3}$. We fix $kz_{\rm IR} = 10^{10}, g^2_5 k = 1, \Delta=10$ and $\sigma_0 = 3$, corresponding to $F_a\simeq 10^9$\,GeV. The curves and bands depict the mean and standard deviation of $\log_{10} F^V$ obtained from a scan over anarchic 5D Yukawa couplings. }
    \label{fig:FV}
\end{figure}

Representative values of the diagonal axial-vector couplings $c_{u,d,\ell}^A$ are shown in figure~\ref{fig:cv-diag} for the quarks and charged leptons. The axion parameters correspond\footnote{For larger values of $F_a\gtrsim 10^{12}$\,GeV that may be required for the axion to account for all of the dark matter, the axion-fermion couplings are suppressed by another factor of $\gtrsim 10^3$ compared to the values shown in figure~\ref{fig:FV}.} to $F_a\simeq 10^9$\,GeV and we take $\Delta=10$ to solve the axion-quality problem. In particular, we see that the top-quark coupling $c_{tt}^A$ is suppressed since it is mostly localized near the UV brane.

\begin{figure}[t]
    \centering
    \includegraphics[width=0.47\textwidth]{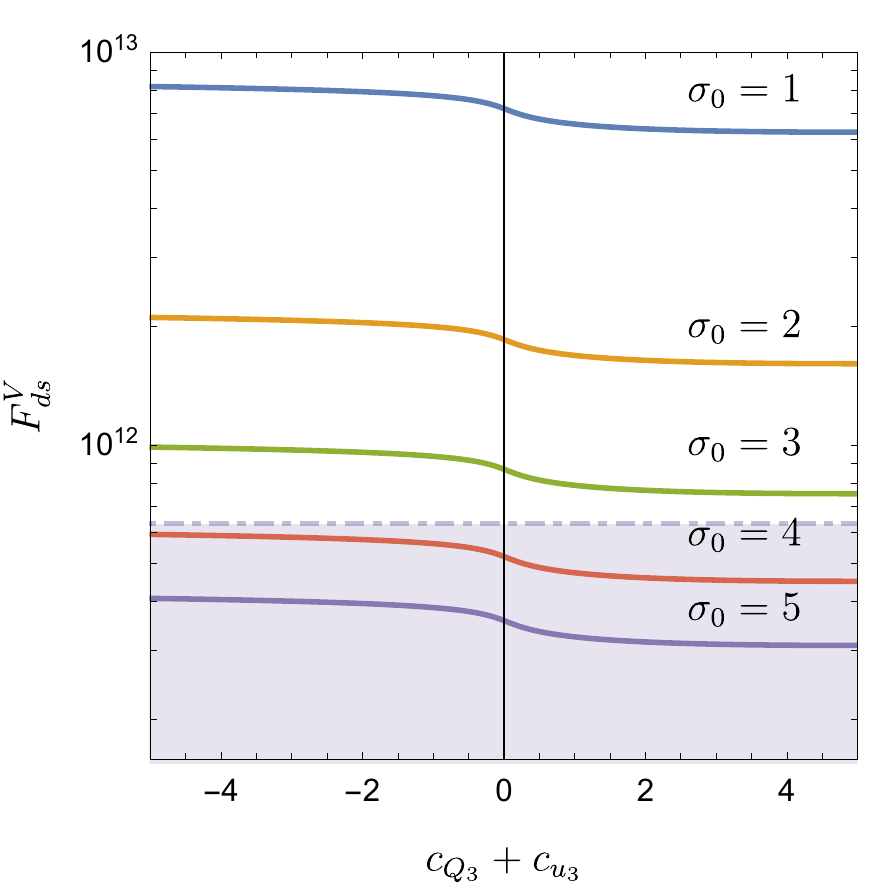} 
    \includegraphics[width=0.47\textwidth]{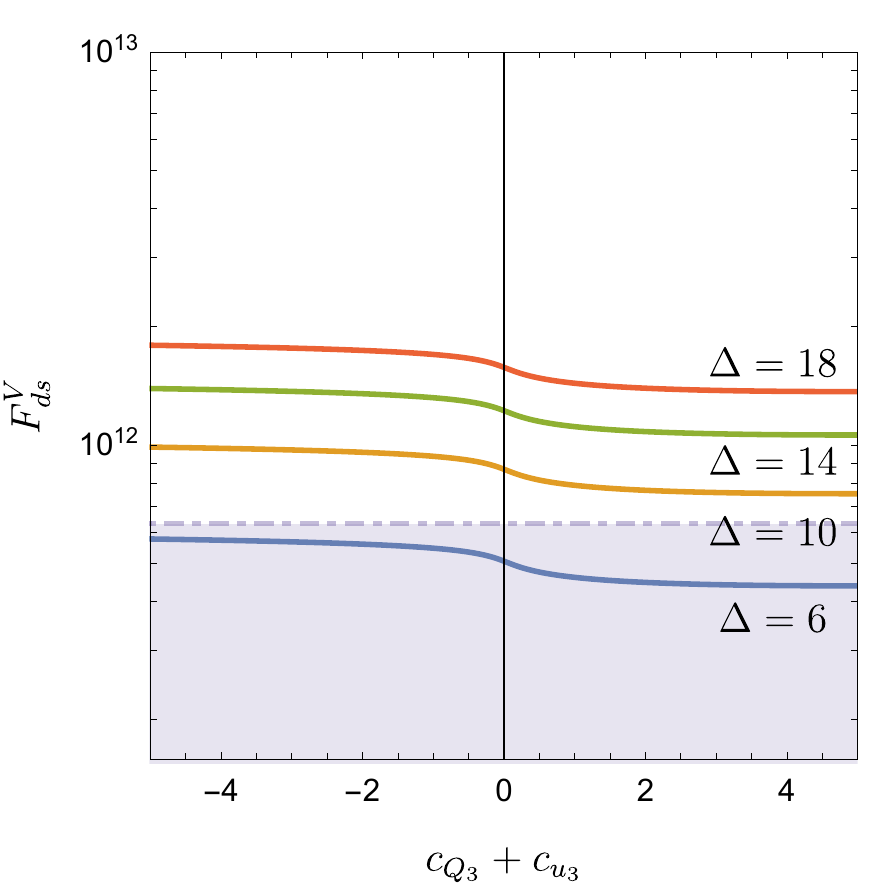} 
    \caption{Absolute values of the axion coupling $F^V_{ds}$ as a function of $c_{Q_3} + c_{u_3}$ for $F_a = 10^9$\,GeV,  $g^2_5 k = 1$ for various values of $\sigma_0$ with $\Delta = 10$ (left), and various values of $\Delta$ with $\sigma_0 = 3$ (right). The region below the dot-dashed line corresponds to the current experimental limit $F_{ds}^V\gtrsim 6.8\times 10^{11}$\,GeV~\cite{MartinCamalich:2020dfe}.}
    \label{fig:sigma_constraint}
\end{figure}

For the off-diagonal couplings, consider first the axion-quark couplings shown in the left panel of figure~\ref{fig:FV} as a function of $c_{Q_3}+c_{u_3}$. 
The $F_u^V$ off-diagonal matrix elements range from approximately $10^{12}-10^{14}$\,GeV, while the $F_d^V$ off-diagonal matrix elements are approximately $10^{11}-10^{13}$\,GeV. The axial-vector couplings $F_{u,d}^A$ are the same order of magnitude as the $F_{u,d}^V$ and are not explicitly shown. These values are comparable to the experimental limits given in Ref.~\cite{MartinCamalich:2020dfe}. Currently, the most stringent limit is $(F_d^V)_{12}\gtrsim 6.8\times 10^{11}$\,GeV from $K^+\rightarrow \pi^+ a$ decays. As shown in figure~\ref{fig:sigma_constraint}, this bound on  $(F_d^V)_{12}$ rules out values of $\sigma_0\gtrsim 4$ for $g_5^2k=1$ and $\Delta=10$. The projected future sensitivity of NA62 and KOTO to $(F_{d }^V)_{12}$ is $2\times 10^{12}$\,GeV~\cite{MartinCamalich:2020dfe}, which can probe values of $\sigma_0\gtrsim 2$.

The axion-charged lepton couplings are obtained in a similar fashion; however, the mixing matrices $A_{L,R}^{e,\nu}$ are sensitive to the mechanism for neutrino masses. For simplicity, we assume that the PMNS matrix is generated in the charged lepton sector ($U_{PMNS}=(A_L^e)^\dagger$) and leave a detailed study of the neutrino sector for future work.  The charged lepton bulk mass parameters are then constrained by fitting the charged lepton masses (see figure~\ref{fig:masses-bulk}) and the PMNS mixing parameters. The resulting $F_l^V$ are shown in the right panel of figure~\ref{fig:FV} as a function of $c_{L_3}+c_{e_3}$, again with $F_a\simeq 10^9$\,GeV and $\Delta=10$. The off-diagonal $F_\ell^V$ matrix elements range from approximately $10^{11}-10^{12}$\,GeV. Again, the axial-vector couplings $F_\ell^A$ are of the same order of magnitude as the vector couplings. The corresponding experimental limits are given in \cite{Calibbi:2020jvd}. The most stringent limit is from $\mu\rightarrow e\, a$, which constrains $(F_{e }^V)_{12} \gtrsim 4.8\times 10^9$\,GeV. Future sensitivity of the MEG-II-fwd and Mu3e experiments is $(F_{e }^V)_{12} \gtrsim 2\times 10^{10}$\,GeV. This is still an order of magnitude smaller than the predicted values shown in figure~\ref{fig:FV}.

%


\subsection{Comparison with 4D flavour axion models}

In 4D flavour models of the Froggatt-Nielsen type \cite{Froggatt:1978nt} where the axion is the phase of the flavon field responsible for generating fermion mass hierarchies,
there are also flavour off-diagonal axion-fermion couplings~\cite{Wilczek:1982rv, Ema:2016ops, Calibbi:2016hwq,Bonnefoy:2019lsn}. Indeed, in such models the Yukawa couplings are typically generated by higher-dimensional operators of the type
\begin{equation}
    \left( {\frac{f}{M}} e^{\frac{ia}{f}}\right)^{X_{Q_i} - X_{u_j} - X_{H_u}} {\bar Q}_{iL} u_{jR} H_u + \cdots \ , 
    \label{fn1}
\end{equation}
where $f$ is the vacuum expectation value of the (radial part of the) flavon field, and $M$ the mass scale where such operators are generated (for simplicity, we have normalized the $U(1)$ charge of the flavon to $1$). After diagonalizing the fermion mass matrices one obtains (using a matrix notation)
\begin{equation}
    -i\frac{\partial_\mu a}{f} \ \left(  {\bar u}_L (A_L^u)^{\dagger}
    X_Q A_L^u \gamma^{\mu} u_L +  {\bar u}_R (A_R^u)^{\dagger}
    X_u A_R^u \gamma^{\mu} u_R + 
    \cdots \ \right)  ,  \label{fn4}
\end{equation}
where the charge matrix $(X_Q)_{ij} = X_{Q_i} \delta_{ij}$, etc. Such off-diagonal couplings are similar to the 5D warped model with bulk Yukawa couplings. The 5D model corresponds to a composite axion which can address the axion quality problem and, depending on the choice of parameters, these couplings may be suppressed relative to $F_a$. This compares with 4D Froggatt-Nielsen models where there is no suppression of such couplings, assuming order one charges. For the massive axion case, as mentioned earlier, the diagonal axion-fermion couplings can also be parametrically suppressed, especially for fermions strongly localized towards the UV brane such as the top quark.  In the Froggatt-Nielsen 4D models, typically the light fermion generations have larger PQ charges and therefore couple numerically (but not parametrically) stronger than the heavier generations.    


\subsection{Axion-gluon/photon couplings}
\label{sec:axion-gluonphoton}

The axion-gluon couplings arise from several sources in the 5D model. Below we consider each separately. We use the known form of anomalies on orbifolds to derive the axion couplings. The direct computation using the axion couplings to all fermionic modes, including the KK modes, is presented in appendix \ref{KKcomputationAppendix}.

\subsubsection{5D anomaly}\label{axionCouplingsSection}

The bulk fermions charged under the U(1)$_{PQ}$ symmetry give rise to 5D gauge anomalies that need to be cancelled. In general, using the results in \cite{ArkaniHamed:2001is,Hirayama:2003kk}, the 5D anomaly is equally distributed on the two boundaries and is given by
\begin{equation} \label{eq:5Danomeqn}
    \eta^{MN} \partial_M J^a_N = \frac{1}{64\pi^2} A(R) d^{abc}\,\epsilon^{\mu\nu\rho\sigma}F^{b}_{\mu\nu}F^{c}_{\rho\sigma}(\delta(z-z_{UV})+\delta(z-z_{IR})) \,,
\end{equation}
where $A(R) d^{abc}\equiv \frac{1}{2} {\rm Tr}(T_{R}^a\{T_{R}^b,T_{R}^c\})$ and $T_R^a$ are the generators for the representation $R$ with normalization ${\rm Tr}(T_F^a T_F^b) = \frac{1}{2}\delta^{ab}$ for the fundamental representation. The U(1)$_{PQ}$ gauge boson boundary conditions are such that the gauge symmetry is restricted to a global symmetry on the UV boundary; hence, only the IR boundary part of the anomaly in \cref{eq:5Danomeqn} needs to be cancelled. 

Consider the $U(1)_{PQ} U(1)_{EM}^2$ and $U(1)_{PQ} SU(3)_c^2$ gauge anomalies. For the electromagnetic anomaly, we obtain $A(R) d_{EM}^{abc} F_{\mu\nu}^b F_{\rho\sigma}^c \equiv A_{EM} F_{\mu\nu} F_{\rho\sigma}$ with a coefficient
\begin{equation}
    A_{EM}=3\left(-\frac{5}{3} X_Q + \frac{4}{3} X_U + \frac{1}{3} X_D -X_L + X_E\right) = -4 \left(X_{H_u} + X_{H_d} \right) \,.
    \label{eq:EManom}
\end{equation}
For the QCD anomaly, $A(R) d_{QCD}^{abc} G_{\mu\nu}^b G_{\rho\sigma}^c \equiv A_{QCD} G_{\mu\nu}^b G_{\rho\sigma}^b$, with $a$ the $U(1)_{PQ}$ index and $b,c$ the $SU(3)_c$ indices,  and the coefficient is
\begin{equation}
    A_{QCD}=\frac{3}{2}(-2 X_Q + X_U + X_D) = - \frac{3}{2} \left(X_{H_u} + X_{H_d} \right) \,.
    \label{eq:QCDanom}
\end{equation}
Since $a(x^\mu,z)$ transforms with a shift under $U(1)_{PQ}$, the IR anomalies in \eqref{eq:5Danomeqn}, with coefficients given by \eqref{eq:EManom} and \eqref{eq:QCDanom}, can be cancelled by introducing the boundary terms
\begin{equation} \label{eq:IRterms}
    -\frac{1}{32\pi^2}A_{EM}\int d^4x\,a \, F \widetilde{F} \bigg|_{z_{IR}} - \frac{1}{32\pi^2}A_{QCD}\int d^4x\,a \, G^c \widetilde{G}^c \bigg|_{z_{IR}}\,,
\end{equation}
where $\widetilde{F}^{\mu\nu}=\frac{1}{2}\epsilon^{\mu\nu\rho\sigma} F_{\rho\sigma}$, and similarly for $\widetilde{G}$. The other anomalies can be similarly cancelled.

The boundary terms in \eqref{eq:IRterms} are not the only source of axion-gluon and axion-photon couplings. There is also a contribution coming from the transformation of the path integral measure under \eqref{eq:5Dfermionredef}. For the photon coupling this is
\begin{equation} \label{eq:jacobian-photon}
    -2 \int_{z_{UV}}^{z_{IR}} d^5x\, 4\left(\frac{a_u}{v_u} + \frac{a_d}{v_d} \right) (\delta(z-z_{UV})+\delta(z-z_{IR})) \frac{1}{32\pi^2} F \widetilde{F} \,,
\end{equation}
and for the gluon coupling
\begin{equation} \label{eq:jacobian-gluon}
    -2 \int_{z_{UV}}^{z_{IR}} d^5x\, \frac{3}{2} \left(\frac{a_u}{v_u} + \frac{a_d}{v_d} \right) (\delta(z-z_{UV})+\delta(z-z_{IR})) \frac{1}{32\pi^2} G^c \widetilde{G}^c \,.
\end{equation}
In addition, there could be one-loop contributions from the KK fermions. In appendix \ref{KKcomputationAppendix} we show that these contributions vanish. 

Combining the terms in \cref{eq:IRterms,eq:jacobian-gluon,eq:jacobian-photon} and using \eqref{eq:bulksub} with $\widehat{f}_{a_X}^0(z)=f_a^0(z)$ for a massless axion gives\footnote{Note that the delta functions are defined such that $2\int_{z_{UV}}^{z_{IR}}\delta(z-z_0)f(z)=f(z_0)$ for $z_0=z_{UV}$ or $z_0=z_{IR}$.}
\begin{equation}
    2 \int_{z_{UV}}^{z_{IR}} d^5x\, f_a^0(z) a^0 \delta(z-z_{UV}) \left( \frac{1}{32\pi^2} A_{EM} F \widetilde{F} + \frac{1}{32\pi^2} A_{QCD} G^c \widetilde{G}^c \right) \,.
\end{equation}
The same expression approximately holds in the massive axion case but with opposite sign, since $\widehat{f}_{a_X}^0(z) \approx f_a^0(z) - f_a^0(z_{UV})$. Finally, integrating over the extra dimension, the couplings in the 4D effective theory are
\begin{equation} \label{eq:EM-QCDcoupling}
    S_{4D} \supset \int d^4x \, a^0 f_a^0(z_{UV}) \left( \frac{e^2}{32\pi^2} A_{EM} F^0 \widetilde{F}^0 + \frac{g_s^2}{32\pi^2} A_{QCD} G^{0c} \widetilde{G}^{0c} \right) \,,
\end{equation}
where $e$ is the electromagnetic coupling, and $g_s$ the QCD coupling. The ratio of the electromagnetic to QCD couplings (after absorbing the gauge couplings into the gauge fields) is then
\begin{equation}
    \frac{E}{N} = \frac{A_{EM}}{A_{QCD}} = \frac{8}{3} \,,
\end{equation}
as obtained in the DFSZ model~\cite{Zhitnitsky:1980tq,Dine:1981rt}. 

The complete 4D effective action for the axion couplings is given by \cref{eq:EM-QCDcoupling,eq:cAcV-bulk}.


\subsubsection{5D Chern-Simons}

An axion-gluon coupling can also be generated via the addition of a 5D Chern-Simons term
\begin{equation}\label{eq:Chern-Simons}
  S_{CS} = -\frac{\kappa}{128\pi^2} \int_{z_{UV}}^{z_{IR}} d^5x\, \epsilon^{MNPQR} V_M G^c_{NP} G^c_{QR} + \frac{\kappa}{64\pi^2}\int d^4x\,a \, G^c \widetilde{G}^c \bigg|_{z_{IR}} \,,
\end{equation}
where $\kappa$ is a dimensionless constant and $\epsilon^{MNPQR}$ is the Levi-Civita tensor density. The second term is needed to cancel the localized gauge anomaly on the IR brane~\cite{Cox:2019rro}. (See appendix \ref{CStermAppendix} for a discussion of the relation between this action and charged heavy bulk fermions.) The 4D effective action then contains the coupling
\begin{equation}
  S_{4D} \supset \int d^4x\, \( f_a^0(z_{IR}) - \int_{z_{UV}}^{z_{IR}} dz\, f_{V_z}^0 \) \frac{\kappa g_s^2}{64\pi^2} a^0 G^{c0} \widetilde{G}^{c0} \,,
  \label{eq:CSaxiongluon}
\end{equation}
and one can show that, regardless of whether there is explicit PQ breaking in the UV, the term in brackets is approximately equal to $z_{IR}\sqrt{\Delta-1}/\sigma_0$. A Chern-Simons term can also be added for the electromagnetic field. 

Finally, note that the axion-gluon/photon couplings in \eqref{eq:EM-QCDcoupling} and  \eqref{eq:CSaxiongluon} are generated at the scale $z_{IR}^{-1}$ which is related to $F_a$. Eventually, far below this scale, the nonperturbative QCD dynamics will generate the usual axion mass from the topological susceptibility. This QCD mass will dominate any mass contribution from explicit violations of the global symmetry on the UV boundary, provided $\Delta\gtrsim 10$~\cite{Cox:2019rro}.


\section{Conclusion}
\label{sec:conclusion}

In this paper we have extended the original DFSZ model to a slice of AdS$_5$, where the axion and Standard Model fermions propagate in the bulk containing a U(1)$_{PQ}$ symmetry. The Higgs fields can be either localized on the UV boundary or propagate in the bulk. The PQ symmetry is spontaneously broken in the bulk and on the IR boundary, while explicit violations of the symmetry are confined to the UV boundary. By choosing large values of $\Delta$, the axion profile becomes sufficiently suppressed at the UV boundary and is therefore insensitive to the explicit violation. This allows the 5D model to naturally incorporate a solution to the axion quality problem~\cite{Cox:2019rro}, and the overlap between the fermion and Higgs profiles can explain
the Standard Model Yukawa coupling hierarchy and mixings. 

The axion-fermion couplings arise from the wavefunction overlap between the axion fields and bulk fermion zero modes, and depend on the localization of the Higgs fields. When the Higgs fields are localized on the UV boundary, only flavour diagonal axion-fermion couplings are obtained due to the orthogonality of the fermion profiles. When the Higgs fields instead propagate in the bulk with a constant VEV, the wavefunction overlap between the $z$-dependent axion profile and the fermion zero modes produces flavour-dependent, off-diagonal axion-fermion couplings. Assuming an axion decay constant $F_a\simeq10^9$\,GeV, the off-diagonal couplings $(F_{u,d,\ell}^V)_{ij}$ range from $10^{11-15}$\,GeV, where the 5D parameters are chosen to solve the axion quality problem and obtain the Standard Model fermion masses and mixings. The off-diagonal axion-fermion couplings are most stringently constrained in the down-quark sector, where the current experimental limit on $F_{sd}^V$~\cite{MartinCamalich:2020dfe} restricts some 5D parameters in the model. Future planned experiments will be able to probe more of the parameter space. The couplings in the lepton sector are less constrained and remain an order of magnitude below future sensitivity. Finally, our model could be generalized by considering $z$-dependent bulk Higgs VEVs that may lead to different predictions for the axion-fermion couplings.

There are also axion-gluon/photon couplings that give rise to $E/N =8/3$, as in the original DFSZ model. These couplings result from a cancellation of the 5D gauge anomalies which are known to be equally split on the two boundaries~\cite{ArkaniHamed:2001is}. Alternatively, as a nontrivial check, the couplings were verified with a direct Kaluza-Klein calculation of triangle Feynman diagrams. Additional contributions to the axion-gluon/photon couplings can be generated from 5D Chern-Simons interactions. These terms can be interpreted as resulting from integrating out extra bulk fermion fields.

By the AdS/CFT correspondence, the 5D model is dual to a 4D composite axion where the strong dynamics has an accidental PQ symmetry with partial compositeness for the Standard Model fermions. The Higgs sector, however, requires a tuning to obtain the electroweak VEV, and therefore possible 4D theories would be similar to those considered in \cite{Gherghetta:2020ofz}. Finally, our setup not only applies to the QCD axion, but can be used to determine the couplings for any axion-like particle. For instance, having a much lower axion decay constant could lead to more stringent constraints on a particular model. Indeed, our results provide further motivation to search for axions via their couplings to fermions.


\section*{Acknowledgements}
M.N. thanks Andrew Miller for useful discussions. The work of T.G. and M.N. is supported in part by the Department of Energy under Grant DE-SC0011842 at the University of Minnesota, and T.G. is also supported by the Simons Foundation. T.G. also thanks the Ecole Polytechnique for hospitality and support where part of this work was done. The work of P.C. is supported by the Australian Government through the Australian Research Council. Q.B. is supported by the Deutsche Forschungsgemeinschaft under Germany's Excellence Strategy  EXC 2121 ``Quantum Universe'' - 390833306 and  E.D. is  supported in part by the ANR grant Black-dS-String ANR-16-CE31-0004-01. This work was also supported in part by the France-US PICS MicroDark.


\clearpage

\appendix

\section{Standard Model fermion masses and mixings}
\label{app:SMflavour}

In this appendix, we describe the masses and mixing of the zero mode fermions and our procedure for fitting the bulk mass parameters, $c_i$, to the measured quark and charged lepton masses and the CKM and PMNS matrices. The 4D Yukawa coupling hierarchy is generated from the overlap of the bulk fermion profiles, assuming order one or ``anarchic'' 5D Yukawa couplings~\cite{Grossman:1999ra,Gherghetta:2000qt}.

Starting with the boundary Higgs case, and taking the up-type quark sector as an example, the fermion mass matrix in the 4D effective theory is given by
\begin{equation}
    m_u^{ij} = y_{u,ij}^{(5)} \frac{v_u}{\sqrt{2}k} f_{Q_{iL}}^0(z_{UV}) f_{U_{jR}}^0(z_{UV}) \,.
\end{equation}
With a bulk Higgs, this generalises to
\begin{equation}
 m_u^{ij} = y_{u,ij}^{(5)} \frac{\sqrt{2}v_u}{\sqrt{k}} \int_{z_{UV}}^{z_{IR}} \frac{dz}{(kz)^5}~f_{Q_{iL}}^0(z) f_{U_{jR}}^0(z)\,.
\end{equation}
Evaluating the overlap integral gives
\begin{align} 
    \int_{z_{UV}}^{z_{IR}} \frac{dz}{(kz)^5}~f_{Q_{iL}}^0(z) f_{U_{jR}}^0(z) &= \frac{\mathcal{N}_{Q_i} \mathcal{N}_{U_j}}{k(c_{Q_i} - c_{U_j})} \( (kz_{UV})^{c_{U_j}-c_{Q_i}} - (kz_{IR})^{c_{U_j}-c_{Q_i}} \) \,, \\
    &\approx \frac{\mathcal{N}_{Q_i} \mathcal{N}_{U_j}}{k(c_{Q_i} - c_{U_j})} (kz_{UV})^
    {c_{U_j}}(kz_{UV})^{-c_{Q_i}}\,, \quad \text{if } c_{Q_i} > c_{U_i} \,. \label{eq:fermion-overlap}
\end{align}
The approximation in the second line holds when $c_{Q_i} > c_{U_i}$, provided $kz_{IR} \gg 1$. This will always be the case we are interested in, since for $c_{Q_i} < c_{U_i}$ the overlap integral is suppressed by $(kz_{IR})^{-n}$, with $n\geq1$; the elements of $m_{ij}$ are then too small to explain the observed quark masses and CKM mixing angles. Using \eqref{eq:fermion-overlap}, the bulk Higgs case can be written in a similar form to the boundary case 
\begin{equation}
 m_u^{ij} = \y[ij]{u} \frac{v_u}{\sqrt{2k}} \tilde{f}_{Q_{i}} \tilde{f}_{U_{j}}\,,
\end{equation}
where 
\begin{equation} \label{eq:effective-profiles}
    \tilde{f}_{Q_{i}} = \frac{\mathcal{N}_{Q_i}}{\sqrt{k}} (kz_{UV})^{-c_{Q_i}} \,, \qquad \tilde{f}_{U_{j}} = \frac{\mathcal{N}_{U_i}}{\sqrt{k}} (kz_{UV})^{c_{U_i}} \,,
\end{equation} 
and $\y[ij]{u} = 2y_{u,ij}^{(5)}/(c_{Q_i} - c_{U_j})$. Note that the $\tilde{f}_i$ are dimensionless, and we have dropped explicit chirality indices for conciseness. There are analogous expressions for the down-type quarks and charged leptons. Below, we focus on the bulk Higgs case; the corresponding results for a boundary Higgs are obtained via the replacement $\tilde{f}_i \to f_i^0(z_{UV})$ and $\tilde{y}^{(5)} \to y^{(5)}/\sqrt{k}$.

The mass matrices are diagonalised via the singular value decomposition $A_L^{u\dagger} m_u^{ij} A_R^u = m_{u_i}$, where $A_{L,R}^u$ are unitary matrices and $m_{u_i}$ is the diagonal matrix containing the up-type masses. To simplify the analysis, we take advantage of the fact that the quark and charged lepton masses are hierarchical and use the approximation scheme of Ref.~\cite{vonGersdorff:2019gle} (with $n=\frac{1}{2}$) for the matrices $A_{L,R}$.

\subsection{Quark sector} 
\label{app:quark-mixing}

We begin with the quark sector, and assume the following scaling for the $\tilde{f}$:
\begin{equation} \label{eq:quark-profile-hierarchy}
    \tilde{f}_{Q_1} \sim \epsilon \tilde{f}_{Q_2} \sim \epsilon^2  \tilde{f}_{Q_3} \,, \qquad \tilde{f}_{U_1} \sim \epsilon \tilde{f}_{U_2} \sim \epsilon^2 \tilde{f}_{U_3} \,, \qquad \tilde{f}_{D_1} \sim \epsilon \tilde{f}_{D_2} \sim \epsilon^2 \tilde{f}_{D_3} \,,
\end{equation}
with $\epsilon \ll 1$. As will become clear from the expressions below, this scaling gives the correct structure to explain the quark masses and CKM elements. Although this additional assumption is not strictly required, it allows us to analytically solve for the $\tilde{f}$ in terms of the masses and CKM elements. This scaling behaviour was also used in Ref.~\cite{Casagrande:2008hr}, where they considered a single, boundary-localized Higgs doublet. To leading order in $\epsilon$, the quark masses are given by the expressions
\begin{align}
    m_u &\simeq \frac{v_u}{\sqrt{2k}} \frac{ | \det \y{u} | }{ | \my[11]{u} | } \tilde{f}_{Q_1} \tilde{f}_{U_1}  \,, \quad 
    &m_d &\simeq \frac{v_d}{\sqrt{2k}} \frac{ | \det \y{d} | }{ | \my[11]{d} | } \tilde{f}_{Q_1} \tilde{f}_{D_1}  \,, \notag \\
    m_c &\simeq \frac{v_u}{\sqrt{2k}} \frac{ | \my[11]{u} | }{ | \y[33]{u} | } \tilde{f}_{Q_2} \tilde{f}_{U_2}  \,, \quad 
    &m_s &\simeq \frac{v_d}{\sqrt{2k}} \frac{ | \my[11]{d} | }{ |\y[33]{d} | } \tilde{f}_{Q_2} \tilde{f}_{D_2}  \,, \notag \\
    m_t &\simeq \frac{v_u}{\sqrt{2k}} | \y[33]{u} | \, \tilde{f}_{Q_3} \tilde{f}_{U_3}  \,, \quad 
    &m_b &\simeq \frac{v_d}{\sqrt{2k}} | \y[33]{d} | \, \tilde{f}_{Q_3} \tilde{f}_{D_3}  \,, 
    \label{eq:app_masses}
\end{align}
where $\my[ij]{}$ denotes the $ij$ minor of the matrix $\y{}$. Using the approximation of \cite{vonGersdorff:2019gle}, and again working to leading order in $\epsilon$, the $A^{u, d}_{L, R}$ matrices are given by
\begin{align}
    A^q_L &\simeq 
    \begin{pmatrix} 
        1 & \frac{ \my[21]{q} }{ \my[11]{q} } \frac{\tilde{f}_{Q_1}}{\tilde{f}_{Q_2}} & 
            \frac{ \y[13]{q} }{ \y[33]{q} } \frac{\tilde{f}_{Q_1}}{\tilde{f}_{Q_3}} \\ 
        - \frac{ \myC[21]{q} }{ \myC[11]{q} } \frac{\tilde{f}_{Q_1}}{\tilde{f}_{Q_2}} & 1 & 
        \frac{ \y[23]{q} }{ \y[33]{q} } \frac{\tilde{f}_{Q_2}}{\tilde{f}_{Q_3}} \\ 
        \frac{ \myC[31]{q} }{ \myC[11]{q} } \frac{\tilde{f}_{Q_1}}{\tilde{f}_{Q_3}} & 
        - \frac{ \yC[23]{q} }{ \yC[33]{q} } \frac{\tilde{f}_{Q_2}}{\tilde{f}_{Q_3}} & 1 
        \end{pmatrix} \,, \qquad \\
    A^q_R &\simeq 
    \begin{pmatrix} 
        1 & \frac{ \myC[12]{q} }{ \myC[11]{q} } \frac{\tilde{f}_{q_1}}{\tilde{f}_{q_2}} &
        \frac{ \yC[31]{q} }{ \yC[33]{q} } \frac{\tilde{f}_{q_1}}{\tilde{f}_{q_3}} \\ 
        - \frac{ \my[12]{q} }{ \my[11]{q} } \frac{\tilde{f}_{q_1}}{\tilde{f}_{q_2}} & 1 & 
        \frac{ \yC[32]{q} }{ \yC[33]{q} } \frac{\tilde{f}_{q_2}}{\tilde{f}_{q_3}} \\ 
        \frac{  \my[13]{q} }{  \my[11]{q} } \frac{\tilde{f}_{q_1}}{\tilde{f}_{q_3}} & 
        - \frac{ \y[32]{q} }{ \y[33]{q} } \frac{\tilde{f}_{q_2}}{\tilde{f}_{q_3}} & 1 
        \end{pmatrix} \cdot
        \text{diag}\( e^{i\phi_1}, e^{i\phi_2}, e^{i\phi_3}\) \,,
    \label{eq:AL_AR_quarks}
\end{align}
with
\begin{equation}
    \phi_1 = \arg(\my[11]{u})-\arg(\det\y{u}) \,, \quad \phi_2 = \arg(\y[33]{u})-\arg(\my[11]{u}) \,, \quad \phi_3 = -\arg(\y[33]{u}) \,.
\end{equation}
Notice that $A_L^q$, and hence the CKM matrix, depends only on ratios of the $\tilde{f}_{Q_i}$ and not on the $\tilde{f}_{u_i}$ or $\tilde{f}_{d_i}$. 

\begin{figure}[ht]
  \centering
    \includegraphics[width=\textwidth]{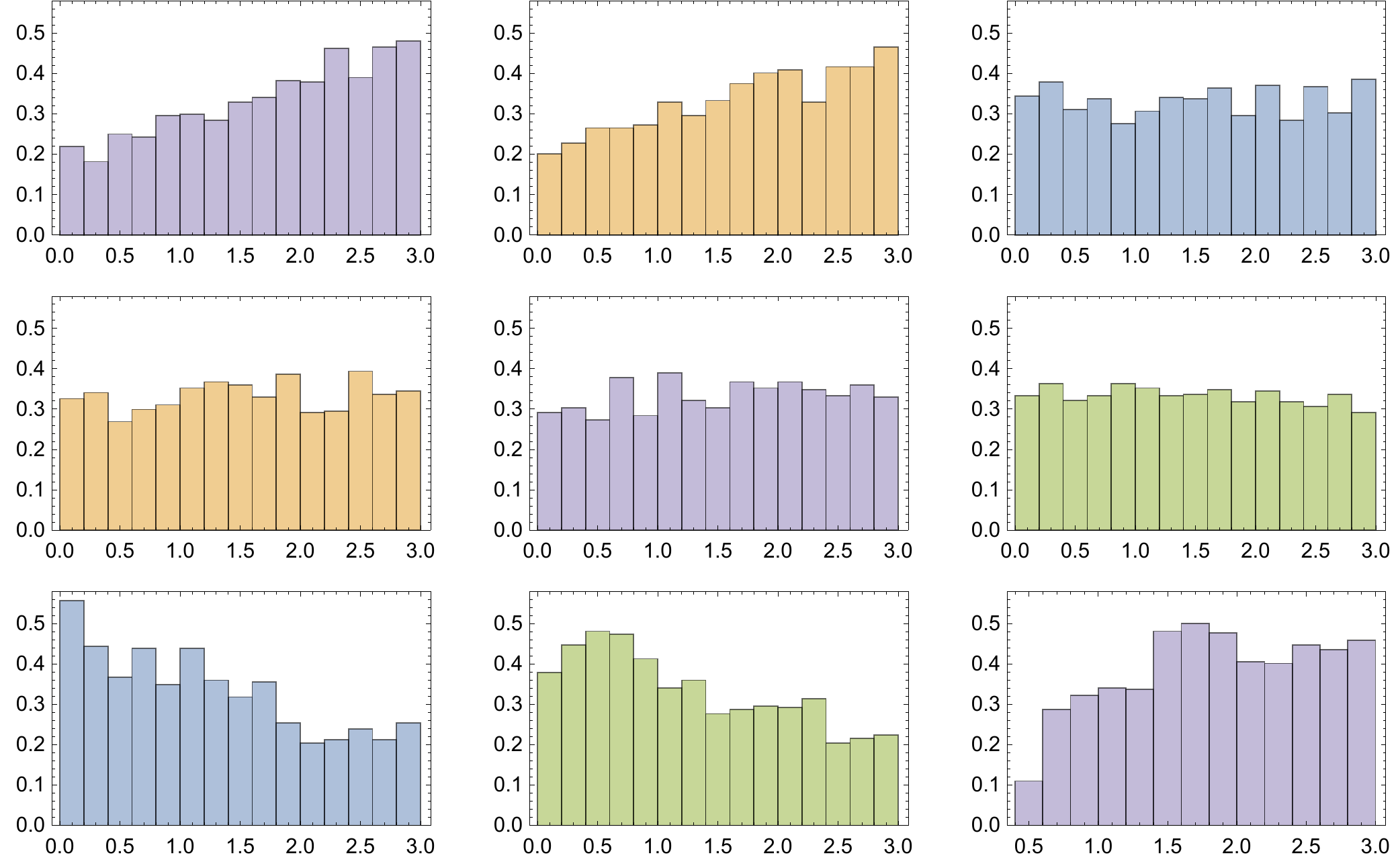}
    \caption{Probability histograms for the absolute values of the randomly generated $\y[ij]{u}$ matrix elements that satisfy~\eqref{eq:rho_eta}.}
    \label{fig:app_1}
\end{figure}

From these expressions one can constrain the 5D Yukawa coupling matrices through the CKM matrix. The Wolfenstein parameters $\bar{\rho}, \bar{\eta}$ are independent of the $\tilde{f}_{Q_i}$ (and hence the quark bulk mass parameters) and are given by
\begin{equation}
    \bar{\rho} - i \bar{\eta} = 
        \frac{  \y[33]{d} \my[31]{u} 
                - \y[23]{d} \my[21]{u} 
                + \y[13]{d} \my[11]{u} }
            { 
                \y[33]{d} \my[11]{u} 
                \left( \frac{\y[23]{d}}{\y[33]{d}} -
                    \frac{\y[23]{u}}{\y[33]{u}} \right) 
                \left( \frac{ \my[21]{d}}{\my[11]{d}} -
                    \frac{\my[21]{u}}{\my[11]{u}} \right) 
            }\,.
    \label{eq:rho_eta}
\end{equation}
We randomly scan over $5 \times 10^6$ samples of 5D Yukawa coupling matrices $\y{u}, \y{d}$ with complex entries of norm between 0 and 3. This results in approximately 1300 matrix pairs that satisfy~\eqref{eq:rho_eta} within $2\,\sigma$ of the experimental values~\cite{Tanabashi:2018oca},
\begin{equation}
    \bar{\rho} = 0.122 \pm 0.018 \,, \qquad 
    \bar{\eta} = 0.355 \pm 0.012 \,.
    \label{eq:CKM_constraint}
\end{equation}
Histograms of the resulting Yukawas, $\y[ij]{u}$, are shown in figure~\ref{fig:app_1}, where approximately uniform distributions can be seen. This is consistent with the assumption that the quark mass hierarchy results from the fermion profiles, rather than the 5D Yukawa couplings. There are similar, although slightly skewed, distributions for $\y{d}$.

\begin{figure}[H]
    \centering
    \includegraphics[width=0.9\textwidth]{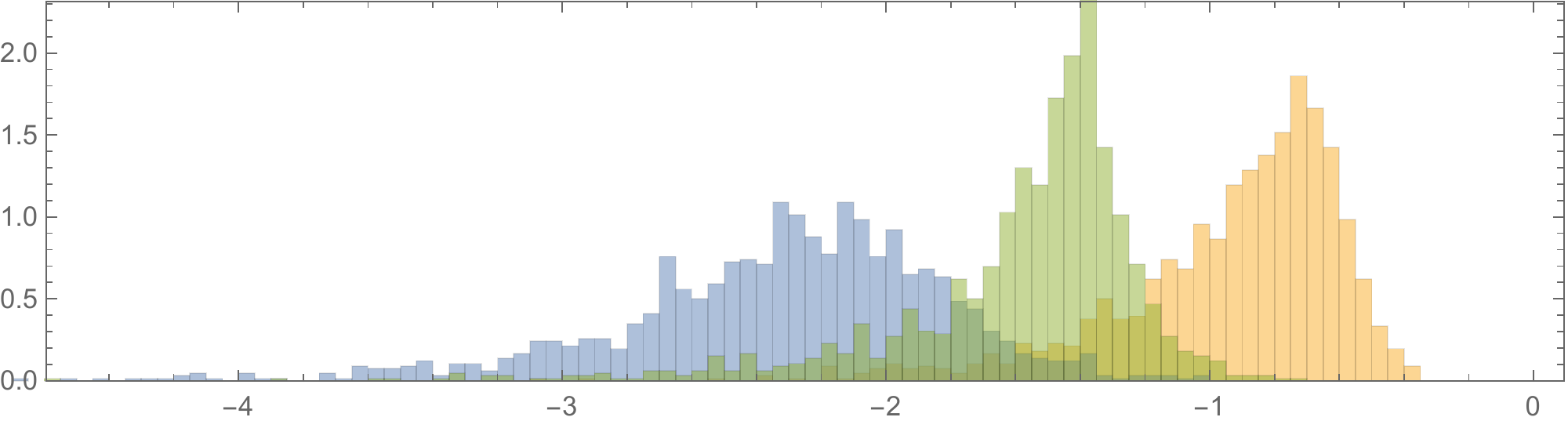} 
    \includegraphics[width=0.9\textwidth]{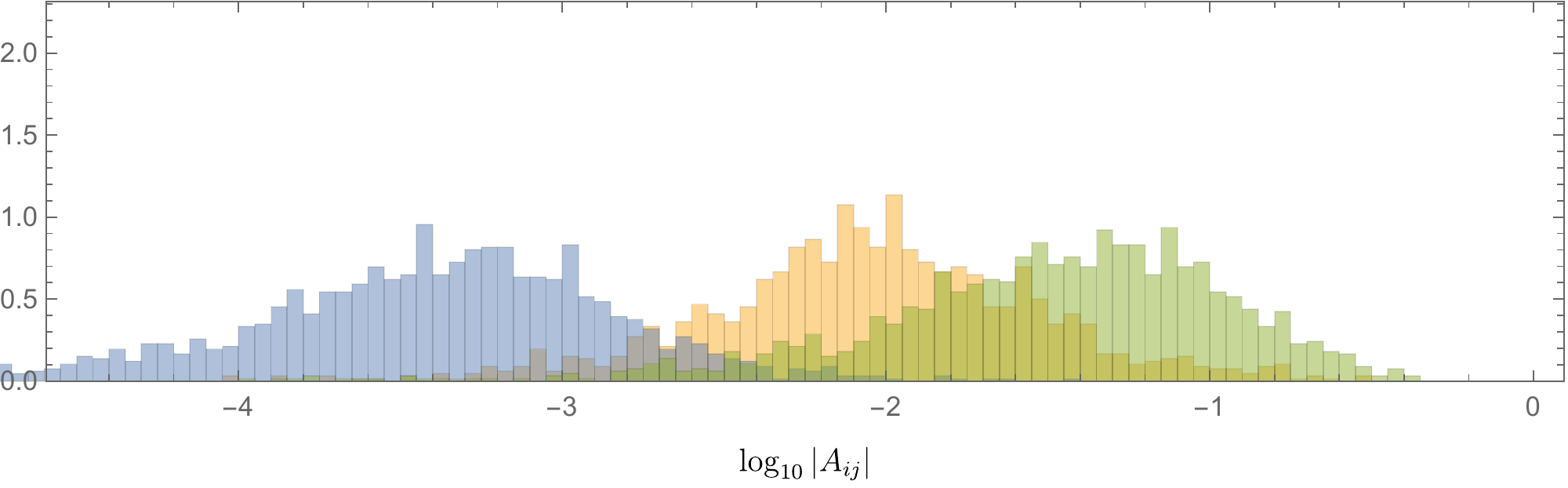}
    \caption{Histograms for the probability distribution of the logarithm (of the absolute value) of the $A^u_L$ (upper) and $A^u_R$ (lower) matrix element 12 (blue), 13 (orange), and 23 (green). 
    }
    \label{fig:hist_Au}
\end{figure}

\begin{figure}[H]
    \centering
    \includegraphics[width=0.9\textwidth]{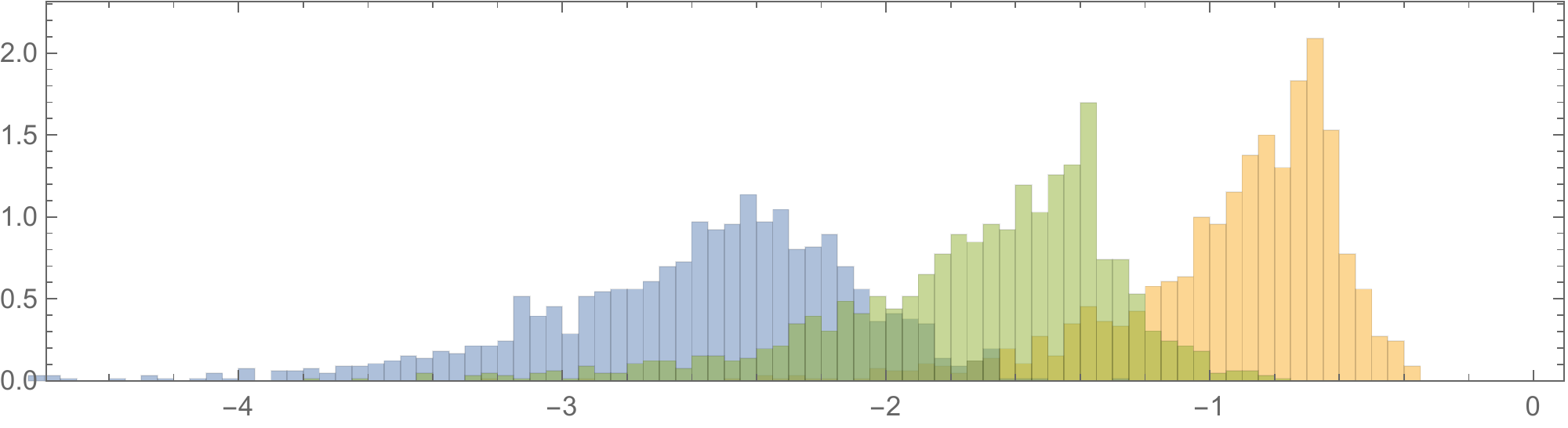} 
    \includegraphics[width=0.9\textwidth]{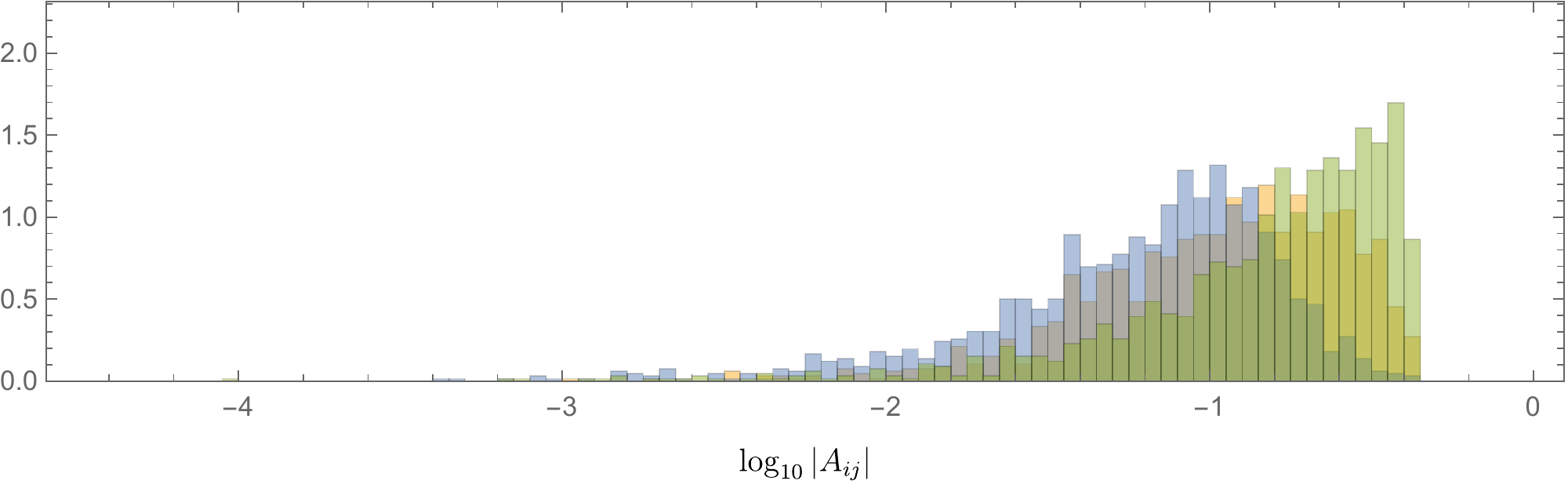}
    \caption{
    Histograms for the probability distribution of the logarithm (of the absolute value) of the $A^d_L$ (upper) and $A^d_R$ (lower) matrix elements 12 (blue), 13 (orange), and 23 (green). 
    }
    \label{fig:hist_Ad}
\end{figure}

Ratios of the $\tilde{f}_{Q_i}$ are constrained through the remaining CKM parameters, $\lambda = 0.22453 \pm 0.00044$ and $A = 0.836 \pm 0.015$, using the expressions
\begin{equation}
    \lambda = 
        \frac{\tilde{f}_{Q_1}}{\tilde{f}_{Q_2}} 
        \left| \frac{\my[21]{d}}{\my[11]{d}} 
            - \frac{\my[21]{u}}{\my[11]{u}} \right|  \,, \quad
    A = \frac{1}{\lambda^2} \frac{\tilde{f}_{Q_2}}{\tilde{f}_{Q_3}} 
        \left| \frac{\y[23]{d}}{\y[33]{d}} - \frac{\y[23]{u}}{\y[33]{u}} \right| \,.
    \label{eq:app_lambda_A}
\end{equation}
Together, \cref{eq:app_masses,eq:app_lambda_A} fix all ratios of the $\tilde{f}_{Q_i}$, $\tilde{f}_{u_i}$ and $\tilde{f}_{d_i}$. The mixing matrices in \cref{eq:AL_AR_quarks} are then fully determined. For each pair of $\y{u}, \y{d}$ that satisfies \eqref{eq:rho_eta} we generate a sample of $A^{u, d}_{L,R}$ matrices. Note that the  mixing matrices in \eqref{eq:AL_AR_quarks} are only approximately unitary; we discard individual matrices that are not unitary to within 20\% accuracy\footnote{We use the $L_1$ norm $||A^{\dagger} A - \mathbb{I}_{3 \times 3}|| \le 0.2$, where the $L_1$ norm is the sum of the absolute values of the matrix elements.}. The resulting distributions for the absolute values of the elements of the $A^u_{L,R}$ and $A^d_{L,R}$ matrices are plotted in figures~\ref{fig:hist_Au} and \ref{fig:hist_Ad}, respectively. The matrices are approximately symmetric, thus only the upper off-diagonal entries are shown. 

Finally, to obtain the axion-fermion couplings from \cref{eq:massiveoverlapint} we need to calculate the overlap integrals of the fermion and axion profiles. Recall that the fermion profiles are related to the $\tilde{f}$ via \cref{eq:effective-profiles}. Since the quark masses and CKM parameters fix only ratios of the $\tilde{f}$, there remains one free parameter which we take to be the combination $c_{Q_3} + c_{U_3}$. Due to the relation $y^{(5)}_{u,ij} = (c_{Q_i} - c_{u_j})\, \tilde{y}^{(5)}_{u,ij}/2$ (and similarly for $y_d^{(5)}$), the bulk mass parameters cannot be too large if the original 5D Yukawa couplings, $y_u^{(5)}$, are to remain perturbative. We find that the largest parameters are $c_{Q_3}$ and $c_{U_3}$ and for $-5 < c_{Q_3} + c_{U_3} < 5$, these always lie in the range $(0.2 , 5.3)$, with the remaining bulk mass parameters between $(-0.1, 3.2)$. The final results for the axion-quark couplings are shown as a function of $c_{Q_3} + c_{U_3}$ in figures~\ref{fig:cv-diag} and~\ref{fig:FV}, where the curves and coloured bands denote the mean and standard deviation of $\log_{10} F^V$ over our sample of ($\y{u}, \y{d}$).

\subsection{Lepton sector}

We can perform a similar analysis for the lepton sector. However, this depends on the precise mechanism for neutrino masses, which can be either Dirac or Majorana and may or may not be hierarchical. For simplicity, we assume that the PMNS matrix is generated in the charged lepton sector. This could follow from a flavour-diagonal Weinberg operator (for instance, by localizing a right-handed neutrino sector on the UV boundary). We leave a more detailed analysis of the neutrino sector possibilities and discussion of axion-neutrino couplings for future work.

An important difference from the quark sector is that the mixing angles in the PMNS matrix are relatively large. As a consequence, an analogous assumption to \eqref{eq:quark-profile-hierarchy} does not work for the charged leptons. Instead, we assume the following scaling
\begin{equation} \label{eq:lepton-texture}
    \tilde{f}_{E_1} \sim \epsilon \tilde{f}_{E_2} \sim \epsilon^2  \tilde{f}_{E_3} \,,
\end{equation}
with $\tilde{f}_{L_i}\sim\epsilon^0$. To leading order in $\epsilon$, the charged lepton masses are then given by
\begin{align} \label{eq:lepton-masses}
    m_e &\simeq \frac{v_d}{\sqrt{2k}} \frac{ | \det \y{e} | }{ N_1 } \frac{\tilde{f}_{L_2}}{\tilde{f}_{L_3}}   \tilde{f}_{L_1} \tilde{f}_{E_1}    \,, \notag \\
    m_{\mu} &\simeq \frac{v_d}{\sqrt{2k}} \frac{N_1}{N_3} \tilde{f}_{L_3} \tilde{f}_{E_2}   \,, \notag \\
    m_{\tau} &\simeq \frac{v_d}{\sqrt{2k}} N_3 \tilde{f}_{L_3} \tilde{f}_{E_3}   \,,
\end{align}
with
\begin{align}
    N_1 &= \sqrt{ |\my[11]{e}|^2 \frac{\tilde{f}_{L_2}^2}{\tilde{f}_{L_3}^2} + |\my[21]{e}|^2 \frac{\tilde{f}_{L_1}^2}{\tilde{f}_{L_3}^2} + |\my[31]{e}|^2 \frac{\tilde{f}_{L_1}^2 \tilde{f}_{L_2}^2}{\tilde{f}_{L_3}^4} } \,, \\
    N_3 &= \sqrt{ |\y[13]{e}|^2 \frac{\tilde{f}_{L_1}^2}{\tilde{f}_{L_3}^2} + |\y[23]{e}|^2 \frac{\tilde{f}_{L_2}^2}{\tilde{f}_{L_3}^2} + |\y[33]{e}|^2 } \,.
\end{align}
The assumption \eqref{eq:lepton-texture} therefore allows us to trivially solve for the $\tilde{f}_{E_i}$, once the $\tilde{f}_{L_i}$ have been obtained from a fit to the PMNS matrix. Again using \eqref{eq:lepton-texture} and working to leading order in $\epsilon$, the mixing matrix $A^e_R$ takes a comparable form to the quark sector: 
\begin{equation} \label{eq:AR_leptons}
    A^e_R \simeq 
    \begin{pmatrix} 
        1 & \frac{ \myC[12]{e} }{ \myC[11]{e} } \frac{\tilde{f}_{E_1}}{\tilde{f}_{E_2}} &
        \frac{ \yC[31]{e} }{ \yC[33]{e} } \frac{\tilde{f}_{E_1}}{\tilde{f}_{E_3}} \\ 
        - \frac{ \my[12]{e} }{ \my[11]{e} } \frac{\tilde{f}_{E_1}}{\tilde{f}_{E_2}} & 1 & 
        \frac{ \yC[32]{e} }{ \yC[33]{q} } \frac{\tilde{f}_{E_2}}{\tilde{f}_{E_3}} \\ 
        \frac{  \my[13]{e} }{  \my[11]{q} } \frac{\tilde{f}_{E_1}}{\tilde{f}_{E_3}} & 
        - \frac{ \y[32]{e} }{ \y[33]{q} } \frac{\tilde{f}_{E_2}}{\tilde{f}_{E_3}} & 1 
        \end{pmatrix} \cdot
        \text{diag}\( e^{-i\arg\det\y{e}}, 1, 1\) \,.
\end{equation}
The expression for $A^e_L$ is significantly more complicated due to the absence of any assumption on the $\tilde{f}_{L_i}$, and is given (to all orders in $\epsilon$) by
\begin{align} \label{eq:AL_leptons}
    A^e_L &\simeq 
    \begin{pmatrix} 
        \frac{\myC[11]{e}}{N_1} \frac{\tilde{f}_{L_2}}{\tilde{f}_{L_3}} & \frac{1}{N_1 N_3} \( \my[21]{e} \yC[33]{e} + \frac{\my[31]{e} \yC[23]{e} \tilde{f}_{L_2}^2}{\tilde{f}_{L_3}^2} \) \frac{\tilde{f}_{L_1}}{\tilde{f}_{L_3}} & \frac{\y[13]{e}}{N_3} \frac{\tilde{f}_{L_1}}{\tilde{f}_{L_3}} \\
        -\frac{\myC[21]{e}}{N_1} \frac{\tilde{f}_{L_1}}{\tilde{f}_{L_3}} & \frac{1}{N_1 N_3} \( \my[11]{e} \yC[33]{e} - \frac{\my[31]{e} \yC[13]{e} \tilde{f}_{L_1}^2}{\tilde{f}_{L_3}^2} \) \frac{\tilde{f}_{L_2}}{\tilde{f}_{L_3}} & \frac{\y[23]{e}}{N_3} \frac{\tilde{f}_{L_2}}{\tilde{f}_{L_3}} \\
        \frac{\myC[31]{e}}{N_1} \frac{\tilde{f}_{L_1}\tilde{f}_{L_2}}{\tilde{f}_{L_3}^2} & \frac{-1}{N_1 N_3} \( \frac{\my[11]{e} \yC[23]{e} \tilde{f}_{L_2}^2}{\tilde{f}_{L_3}^2} + \frac{\my[21]{e} \yC[13]{e} \tilde{f}_{L_1}^2}{\tilde{f}_{L_3}^2} \) & \frac{\y[33]{e}}{N_3} \\
        \end{pmatrix} \,.
\end{align}
Given our assumption regarding the neutrino sector, $A^e_L$ is directly related to the PMNS matrix: $U_{PMNS} =  (A^e_L)^\dagger A^\nu_L = (A^e_L)^\dagger$, since $A^\nu_L$ is the identity matrix. The mixing angles and Dirac phase of the PMNS matrix are then simply
\begin{align}
    \tan\theta_{12} &= \left|\frac{\my[21]{e}}{\my[11]{e}}\right| \frac{\tilde{f}_{L_1}}{\tilde{f}_{L_2}} \,, \label{eq:PMNStheta12} \\
    \tan\theta_{23} &= \frac{1}{N_1 |\y[33]{e}|} \left| \myC[11]{e} \y[23]{e} \frac{\tilde{f}_{L_2}^2}{\tilde{f}_{L_3}^2} + \myC[21]{e} \y[13]{e} \frac{\tilde{f}_{L_1}^2}{\tilde{f}_{L_3}^2} \right| \,, \label{eq:PMNStheta23} \\
    \sin\theta_{13}\,e^{-i\delta_{CP}} &= \frac{\my[31]{e}}{N_1} \frac{\tilde{f}_{L_1}\tilde{f}_{L_2}}{\tilde{f}_{L_3}^2} \,. \label{eq:PMNStheta13}
\end{align}
We use the results from the fit to the oscillation data in~\cite{Esteban:2020cvm}, with normal ordering: $\theta_{12}/^{o} = 33.44^{+ 0.78}_{-0.75}$, $\theta_{23}/^{o} = 49.0^{+ 1.1}_{-1.4}$, $\theta_{13}/^{o} = 8.57^{+ 0.13}_{-0.12}$ and $\delta_{CP}/^{o} = 195^{+ 51}_{-25}$.

\begin{figure}[ht]
    \centering
    \includegraphics[width=0.9\textwidth]{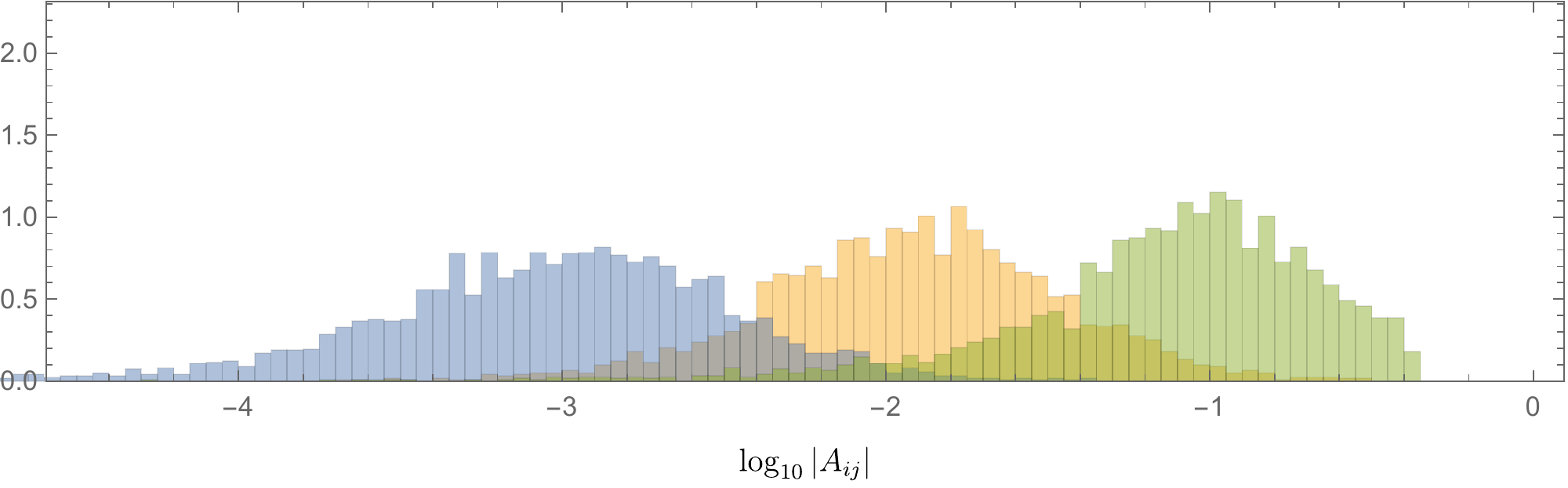}
    \caption{
    Histograms for the probability distribution of the logarithm (of the absolute value) of the $A^e_R$ matrix elements 12 (blue), 13 (orange), 23 (green). $A^e_L$, given by Eq.~\eqref{eq:AL_leptons}, is directly correlated with the PMNS matrix and thus is not plotted.
    }
    \label{fig:hist_Ae}
\end{figure}

We randomly scan over approximately $2 \times 10^5$ possible 5D Yukawa coupling matrices $\y{e}$, with complex entries of norm between 0 and 3. Equations~\eqref{eq:PMNStheta12} and \eqref{eq:PMNStheta13} are then used to solve for the ratios $\tilde{f}_{L_1} / \tilde{f}_{L_3}$ and $\tilde{f}_{L_2} / \tilde{f}_{L_3}$. These ratios are substituted into equation~\eqref{eq:PMNStheta23}, which is required to satisfy the experimental value at $2\,\sigma$. This process yields about 2400 viable matrices $\y{e}$. After imposing the relations \eqref{eq:lepton-masses} for the charged lepton masses, all ratios of the $\tilde{f}_{L_i}$ and $\tilde{f}_{E_i}$ are fixed. The matrix $A_R^e$ is then also determined, and distributions of the absolute values of the elements are plotted in figure~\ref{fig:hist_Ae}. $A^e_R$ is approximately symmetric, and hence only the upper off-diagonal elements are shown. 

Finally, the axion-fermion couplings are calculated as a function of the remaining free parameter, which we take to be the combination $c_{L_3} + c_{E_3}$. The results are shown in figures~\ref{fig:cv-diag} and~\ref{fig:FV}, where the curves and coloured bands denote the mean and standard deviation of $\log_{10} F^V$ over our sample of $\y{e}$.

\section{Direct Kaluza-Klein calculations}\label{KKcomputationAppendix}

In this appendix, we further justify the axion couplings to gauge bosons that are presented in Section \ref{axionCouplingsSection}. We directly compute those interactions from the axion couplings to the fermionic Kaluza-Klein (KK) towers associated with the bulk fermions.

\subsection{Useful formulae}\label{preliminariesAppendix}

We first list a few formulae that we will use later on. 

\paragraph{Axion-gauge bosons EFT couplings} To begin with, we will need the one-loop axion couplings that arise when integrating out a 4D fermion. More precisely, consider the theory of an axion field $a$, and a Dirac fermion $\psi$ in the fundamental representation of a gauge group with generators $T^a$, gauge field $A_\mu^a$ and field strength $F_{\mu\nu}^a$ (we keep the group unspecified for now, and we identify it later with QCD or electromagnetism). The Lagrangian is given by
\begin{equation}
  \cL_{4D} = -\frac{1}{2g^2}\text{Tr}(F_{\mu\nu}^2)-\overline{\psi}\gamma^\mu \mathcal{D}_\mu \psi-m_\psi\overline\psi\psi-\frac{1}{2}(\partial_\mu a)^2+iya\overline{\psi}\gamma_5\psi \ ,
\end{equation}
where $\mathcal{D}_\mu=\partial_\mu-iA_\mu^aT^a$, and in the EFT below the fermion mass, $m_\psi$, the one-loop dimension-5 axion-gauge field coupling reads \cite{Steinberger:1949wx}
\begin{equation}
  \cL_{\text{eff},4D} \supset -\frac{y}{16\pi^2m_\psi}a\text{Tr}(F_{\mu\nu}\widetilde F^{\mu\nu}) \ .
\label{equation:axionViaLoops}
\end{equation}

\paragraph{Kaluza-Klein equations with mixing} We will also use the precise properties of the KK expansion for the bulk fermions in \eqref{KKexpansionBulkFermions}. In order to determine the axion EFT below the mass of all fermions we must resolve the mass of the would-be KK zero-modes, which means that we must take into account the mixing due to the bulk Higgs VEVs.

Consider the theory of two 5D fermions $\psi$ and $\chi$ with a small mixing. The action is
\be
\bead
  S_{5D}=-2\int d^5x\sqrt{-g}\[\frac{1}{2}\(\overline\psi\Gamma^M\mathcal{D}_M\psi-\mathcal{D}_m\overline\psi\Gamma^M\psi\)+M_\psi\overline\psi\psi+(\psi\leftrightarrow\chi)+y\langle\phi\rangle\overline\psi\chi+\overline y\langle\phi^\dagger\rangle\overline\chi\psi\] \,.
\eead
\label{5DlagOneFlavour}
\ee
Henceforth we take $y\langle\phi\rangle$ to be real. Decomposing $\psi=\big(\begin{matrix}\psi_L&\psi_R\end{matrix}\big)^T$ (and similarly for $\chi$), the equations of motion are
\be
\gamma^\mu\partial_\mu\psi_{L(R)}\mp\partial_z\psi_{R(L)}+\(AM_\psi\mp\frac{2A'}{A}\)\psi_{R(L)}+Ay\langle\phi\rangle\chi_{R(L)}=0 \ ,
\label{5Deoms}
\ee
and similarly with $\psi\leftrightarrow\chi$. To cancel the boundary variation, we impose Dirichlet boundary conditions on $\psi_R$ and $\chi_L$. Then, we introduce the 4D KK massive modes $\xi_{L(R)}^n(x^\mu)$ that satisfy $\gamma^\mu\partial_\mu\xi^n_{L(R)}=-m_n\xi^n_{R(L)}$, and we decompose
\be
  \psi_{L(R)}=\sum_nf_{\psi,L(R)}^n(z)\xi^n_{L(R)}(x^\mu) \ , \quad \chi_{L(R)}=\sum_nf_{\chi,L(R)}^n(z)\xi^n_{L(R)}(x^\mu) \ .
\label{KKdecompositions}
\ee
The equations of motion imply that the KK modes satisfy
\be
  -m_nf_{\psi,L(R)}^n\mp\partial_zf_{\psi,R(L)}^n+\(AM_\psi\pm\frac{2A'}{A}\)f_{\psi,R(L)}^n+Ay\langle\phi\rangle f_{\chi,R(L)}^n=0 \ ,
\label{KKeqs}
\ee
and similarly with $\psi\leftrightarrow\chi$. To obtain properly normalized kinetic terms we require that
\be
  2\int dz\, A^4\(f_{\psi,L(R)}^nf_{\psi,L(R)}^m+f_{\chi,L(R)}^nf_{\chi,L(R)}^m\)=\delta^{mn} \ .
\label{orthonormality}
\ee
Finally, the KK modes satisfy a very useful completeness relation\footnote{This is most easily seen by expanding the 5D fermions both in terms of the ``mixed'' KK modes, namely those which satisfy \eqref{KKeqs}, and the ``unmixed'' ones, which satisfy \eqref{KKeqs} when $y=0$. Then, by matching the two expansions, one can export completeness relations for the ``unmixed'' KK modes, found e.g. in \cite{ArkaniHamed:2001is,Hirayama:2003kk}, to the ``mixed'' ones.},
\be
  A^4\sum_n(f_{\psi,L}^n{}^2-f_{\psi,R}^n{}^2)=-A^4\sum_n(f_{\chi,L}^n{}^2-f_{\chi,R}^n{}^2)=\frac{1}{2}\left[\delta(z-z_{UV})+\delta(z-z_{IR})\right]\ .
\label{completenessSumWithMixing}
\ee
The signs in \eqref{completenessSumWithMixing} would be reversed had we chosen Dirichlet boundary conditions for $\psi_L$ and $\chi_R$ instead.

All the above relations can be generalized to the case of several 5D fermion flavours. In this case, the Yukawa $y$ and the KK functions carry flavour indices and are generally complex.

\subsection{Axion couplings to gauge fields}

We now derive the one-loop axion couplings to photons and gluons in the bulk Higgs model of Section \ref{bulkHiggsSection}, using the explicit axion couplings to the KK modes.

Since we focus on photons and gluons, and given the structure of the mixing terms in \eqref{eq:bulkYaction}, we can ignore the full $SU(2)$ structure and simply project each doublet onto its individual components. Therefore, we only present the computations involving the up-type quarks. We also restrict to a single flavour in order to reduce the clutter of indices, with the generalization to three flavours being straightforward. We sum over all SM fermions and all generations at the end.

Writing the quark doublet as $Q=\big(\begin{matrix}Q_u&Q_d\end{matrix}\big)^T$, the action in the up-type sector is
\be
\bead
  S_{5D}=-2\int d^5x\sqrt{-g}\bigg[\frac{1}{2}(\overline Q_u\Gamma^MD_MQ_u-D_M\overline Q_u\Gamma^MQ_u)&+M_Q\overline Q_uQ_u+(Q_u\leftrightarrow U)\\
  &+\frac{y_uv_u}{\sqrt{2}}e^{i\frac{a_u}{v_u}}\overline Q_uU+\frac{y_uv_u}{\sqrt{2}}e^{-i\frac{a_u}{v_u}}\overline UQ_u\bigg] \ .
\eead
\label{actionUsector}
\ee
We introduce the KK modes $U^n$ as in appendix \ref{preliminariesAppendix}:
\be
  Q_{u,L(R)}=\sum_nf_{Q_u,L(R)}^n(z)U_{L(R)}^n(x^\mu) \ , \quad U_{L(R)}=\sum_nf_{U,L(R)}^n(z)U_{L(R)}^n(x^\mu) \ .
\ee
We extract the couplings to $V_z$ from the covariant derivatives, and the $a_u$ couplings from the mixing term. Projecting onto the axion zero modes using the KK expansions \eqref{eq:KK-expansion}, and taking into account that the couplings to gluons and photons do not mix KK modes, \eqref{actionUsector} leads to the following relevant 4D axion couplings,
\be
\bead
  S_{5D}\supset2\int d^4x \, ia^0\bigg[\int dz\,&A^4f_{V_z}^0\(X_Qf_{Q_u,R}^{n}f_{Q_u,L}^{n}+X_uf_{U,R}^{n}f_{U,L}^{n}\)\\
  &-\frac{y_uv_uX_{H_u}}{\sqrt 2N_X}A^5f_{a_X}^0\(f_{Q_u,R}^{n}f_{U,L}^{n}-f_{U,R}^{n}f_{Q_u,L}^{n}\)\bigg]\overline U^n\gamma_5U^n \ .
\eead
\label{relevantUVcouplings}
\ee
We can now use the KK equations \eqref{KKeqs}, which imply that
\be
\bead
  \partial_z(A^4X_Qf_{Q_u,R}^{n}f_{Q_u,L}^{n})&=X_QA^4\[m_n\(f_{Q_u,R}^{n,2}-f_{Q_u,L}^{n,2}\)-\frac{Ay_uv_u}{\sqrt 2}\(f_{Q_u,R}^{n}f_{U,L}^{n}-f_{Q_u,L}^{n}f_{U,R}^{n}\)\] \ ,\\
  \partial_z(A^4X_uf_{U,R}^{n}f_{U,L}^{n})&=X_uA^4\[m_n\(f_{U,R}^{n,2}-f_{U,L}^{n,2}\)+\frac{Ay_uv_u}{\sqrt 2}\(f_{Q_u,R}^{n}f_{U,L}^{n}-f_{Q_u,L}^{n}f_{U,R}^{n}\)\] \ ,
\eead
\ee
as well as the relations $f_{a_X}^0\approx N_X f_a^0$, $\partial_zf_a^0=f_{V_z}^0$, $X_{H_u}=X_Q-X_u$ (see \cref{bulkHiggsSection}) and \eqref{completenessSumWithMixing}, to show that \eqref{relevantUVcouplings} simplifies to
\be
\bead
  S_{5D}\supset\int d^4x \, \frac{i}{2} X_{H_u}m_n\(f_a^0(z_{UV})+f_a^0(z_{IR})\)a^0\overline U^n\gamma_5U^n \ .
\eead
\label{relevantUVcouplings2}
\ee
Using the first part of appendix \ref{preliminariesAppendix}, we obtain the coupling to gluons:
\be
  \cL_{\text{eff},4D} \supset -\frac{1}{64\pi^2}X_{H_u}\(f_a^0(z_{UV})+f_a^0(z_{IR})\)a^0 G^c\widetilde G^c \ .
\ee
The photon coupling is similarly obtained,
\be
  \cL_{\text{eff},4D} \supset  -\frac{1}{24\pi^2}X_{H_u}\(f_a^0(z_{UV})+f_a^0(z_{IR})\)a^0
F\widetilde F \ .
\ee
Summing over the different SM fermions, the complete axion one-loop couplings arising from the bulk fermions become
\be
\bead
  \cL_{\text{eff},4D} \supset &-\frac{3}{64\pi^2}(X_{H_u}+X_{H_d})\(f_a^0(z_{UV})+f_a^0(z_{IR})\)a^0 G^c\widetilde G^c\\
  &-\frac{1}{8\pi^2}(X_{H_u}+X_{H_d})\(f_a^0(z_{UV})+f_a^0(z_{IR})\)a^0F\widetilde F \ ,
\eead
\ee
which is consistent with \eqref{eq:EM-QCDcoupling}. We have explicitly checked, by repeating the same calculation for three flavours using flavoured KK functions, that the mixings between different flavours are irrelevant.

Note that similar computations can be performed to check that after the field redefinition in \eqref{eq:5Dfermionredef}, the couplings of the axion to the bulk fermions do not contribute at one-loop. After the redefinition in \eqref{eq:5Dfermionredef}, which removes the axion from the mixing terms, the axion couplings from the kinetic terms are shifted as follows,
\be
\bead
  &iX_QA^4\overline Q_u\gamma_5 Q_uV_z \longrightarrow iA^4\overline Q_u\gamma_5 Q_u \left(X_QV_z-\beta\frac{\partial_za_u}{v_u}\right) \ ,\\
  &iX_uA^4\overline U\gamma_5 UV_z \longrightarrow iA^4\overline U\gamma_5 U \left(X_uV_z-(\beta-1)\frac{\partial_za_u}{v_u}\right) \ .
\eead
\ee
We ignore the additional couplings that contain 4D derivatives of the axion since they do not play a role in the photon or gluon couplings \cite{Quevillon:2019zrd,Bonnefoy:2020gyh}. The relevant axion-fermion couplings are now
\be
\bead
  S_{5D}\supset\int d^4x \, ia^0\bigg[\int dz\,A^4\bigg(&\left[X_Qf_{V_z}^0-\beta\frac{X_{H_u}}{N_X}\partial_zf_{a_X}^0\right]f_{Q_u,R}^{n}f_{Q_u,L}^{n}\\
  &+\left[X_uf_{V_z}^0-(\beta-1)\frac{X_{H_u}}{N_X}\partial_zf_{a_X}^0\right]f_{U,R}^{n}f_{U,L}^{n}\bigg)\bigg]\overline U^n\gamma_5U^n \ .
\eead
\label{leftOverRedefinition}
\ee
Again using the KK equations, the relations between the different profiles\footnote{Note that the relations between $f_a^0$, $f_{V_z}^0$ and $f_{a_X}^0$ were derived for the massless axion case. In the massive case those relations are not expected to hold exactly, however one can obtain relations using an expansion in powers of $g_5\sqrt k$~\cite{Cox:2019rro} (see also the discussion around Eq.~\eqref{eq:a_X-profile}). Doing so, one sees that all the terms in \eqref{leftOverRedefinition} are of order $g_5^2k$, and the results obtained in the massless case still hold at leading order.}, and the completeness relations for the KK functions, one sees that the one-loop contribution coming from \eqref{leftOverRedefinition} vanishes. The axion couplings are therefore simply given by the path integral transformation due to the chiral redefinition, as anticipated in \eqref{eq:jacobian-photon} and \eqref{eq:jacobian-gluon}.

\subsection{4D UV Peccei--Quinn anomaly}

As a consistency check, one can show that the projection of the 5D anomaly in \eqref{eq:5Danomeqn} on the axion zero mode, or equivalently the result of the KK computation above, can be reinterpreted as the 4D Peccei--Quinn anomaly of the KK theory. Indeed, the 4D global PQ symmetry arises from the 5D PQ gauge symmetry, when the latter is restricted to a local parameter of the form
\be
  \alpha_\text{PQ}(x^\mu,z)=\alpha_0f_a^0(z) \ .
\ee
Focusing again on a single flavour for the up-type quarks, this symmetry generator acts on the fields as
\be
  Q_u\rightarrow e^{i\alpha_\text{PQ}X_Q}Q_u \ , \quad U\rightarrow e^{i\alpha_\text{PQ}X_u}U \ ,
\ee
or, infinitesimally,
\be
  \delta_\text{PQ}Q_u= i\alpha_\text{PQ}X_QQ_u \ , \quad \delta_\text{PQ}U= i\alpha_\text{PQ}X_uU \ .
\ee
Using the orthogonality condition for the KK modes in \eqref{orthonormality}, we obtain the transformation properties of the KK modes,
\be
\bead
  \delta_\text{PQ}U_{L(R)}^n&=\delta_\text{PQ}\(2\int dz\,A^4\[f_{Q_u,L(R)}^nQ_u+f_{U,L(R)}^nU\]\)\\
  &=i\alpha_0\(2\int dz\,A^4f_a^0\[X_Qf_{Q_u,L(R)}^nf_{Q_u,L(R)}^m+X_uf_{U,L(R)}^nf_{U,L(R)}^m\]\)U_{L(R)}^m \ .
\eead
\ee
Therefore, the generator $T_\text{PQ}$ of the PQ symmetry mixes the KK modes,
\be
  T_\text{PQ}^{mn}=2\int dz\,A^4f_a^0\[X_Qf_{Q_u,L(R)}^nf_{Q_u,L(R)}^m+X_uf_{U,L(R)}^nf_{U,L(R)}^m\] \ ,
\ee
and its mixed anomaly coefficient with the QCD gauge group (of generators $T^a$) is
\be
  \[\text{Tr}_L(T_\text{PQ})-\text{Tr}_R(T_\text{PQ})\]\text{Tr}(T^aT^b)=\frac{1}{2}X_{H_u}\(f_a^0(z_{UV})+f_a^0(z_{IR})\)\text{Tr}(T^aT^b) \ ,
\ee
while for the electromagnetic anomaly, we find
\be
  3\times\[\text{Tr}_L(T_\text{PQ})-\text{Tr}_R(T_\text{PQ})\]\(\frac{2}{3}\)^2=\frac{2}{3}X_{H_u}\(f_a^0(z_{UV})+f_a^0(z_{IR})\) \ ,
\ee
consistently with the results obtained above. 

\section{5D Chern-Simons term and boundary axion couplings}\label{CStermAppendix}

In this appendix, we discuss the interpretation of the axion couplings and the 5D anomaly in terms of a 5D Chern-Simons interaction. This will also allow us to show how the effective action \eqref{eq:Chern-Simons} can arise. For definiteness, we consider a bulk fermion $\psi$, with action
\be
  S_{5D}=-2\int d^5x\sqrt{-g}\[\frac{1}{2}(\overline\psi\Gamma^M\mathcal{D}_M\psi-\mathcal{D}_m\overline\psi\Gamma^M\psi)+M_\psi\overline\psi\psi\]+\text{boundary terms} \ ,
\label{5DactionCSUV}
\ee
with $\mathcal{D}_M=\partial_M-i V_M-iA_M^aT^a$, where $V_M$ is the PQ gauge field and $A_M^a$ the gluon gauge field, of field strength $G_{MN}^a$ (the generalization to photons is straightforward). The choice of boundary terms and boundary conditions will be discussed below.

\subsection{Integrating out Kaluza-Klein modes}

In a non-compact 5D space, it is known that the low-energy theory below the mass of a charged fermion contains a non-decoupling Chern-Simons term \cite{Witten:1996qb}. It turns out that the same result can be obtained upon integrating out the KK modes of a bulk fermion \cite{ArkaniHamed:2001is}. We now proceed to show this explicitly for the bulk fermion in \eqref{5DactionCSUV} propagating in a slice of AdS.

The KK decomposition of the 5D fermion is given by
\be
  \psi_{L(R)}(x^M)=\sum_nf_{L(R)}^n(z)\psi_{L(R)}^n(x^\mu) \ ,
\label{KKdecom}
\ee
where the 4D fields $\psi_{L(R)}^n$ have masses $m_n$ and the KK profiles $f_{L(R)}^n(z)$ satisfy the relations presented in \eqref{KKeqs} when the mixing vanishes ($y=0$).

\begin{figure}[h]
  \centering
  \includegraphics[width=0.4\textwidth]{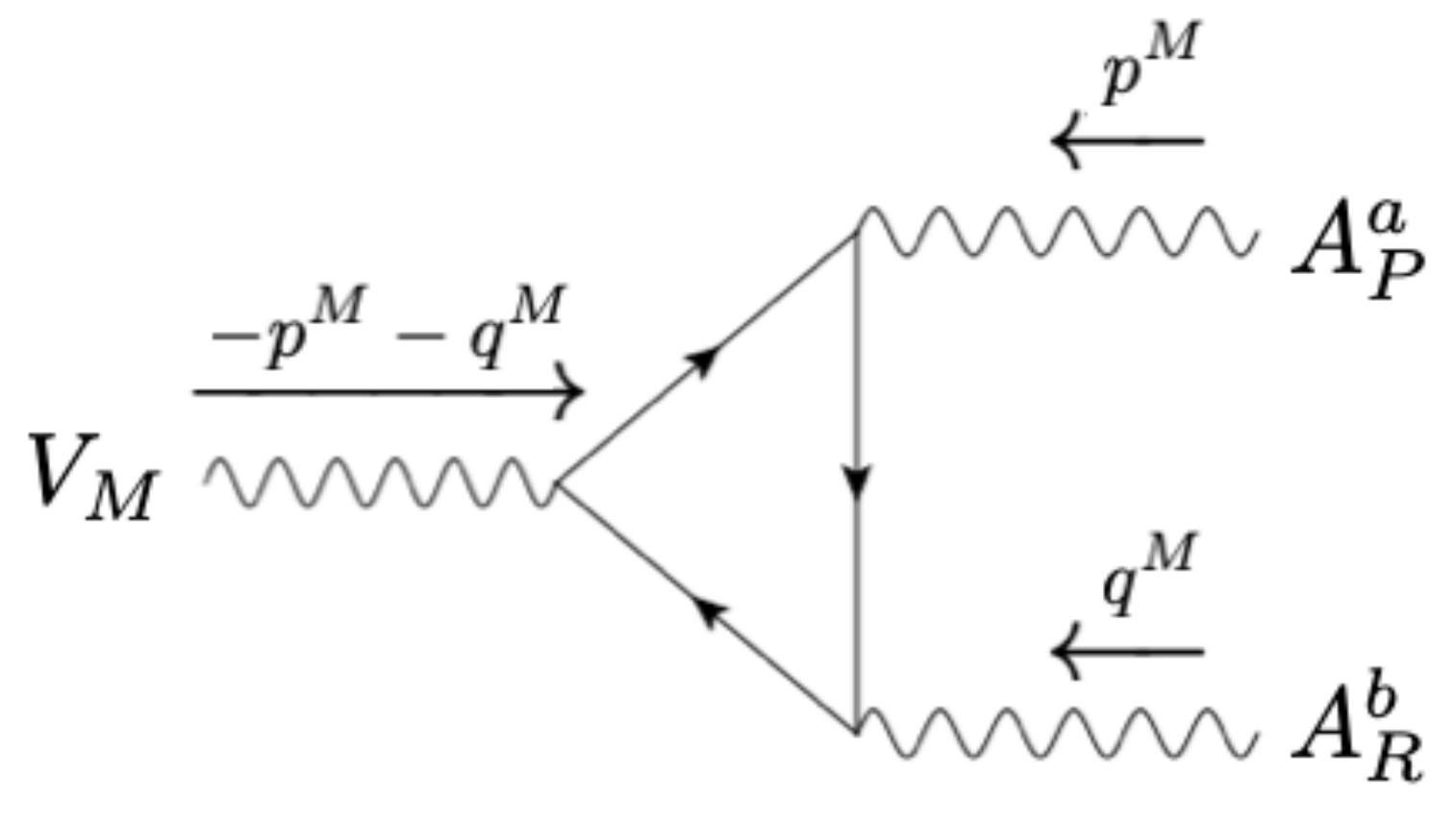}
  \caption{An example of a triangle loop diagram generating the Chern-Simons term}
  \label{CStriangle}
\end{figure}

To identify the effective Chern-Simons term, we compute the 5D Feynman diagrams with bulk fermion propagators. The latter are expressed in terms of the 4D KK fermion propagators (see e.g. \cite{Binetruy:1998hg}). For the chiral components $\psi_{L(R)}$, we obtain
\be
\bead
  \bra{0}T \psi_L(x,z_1)\overline{\psi_L}(y,z_2)\ket{0}=&-i\sum_nf_L^n(z_1)f_L^n(z_2)P_L\int \frac{d^4p}{(2\pi)^4}e^{ip(x-y)}\frac{-i\slashed p}{p^2+m_n^2-i\epsilon} \ , \\
  \bra{0}T \psi_L(x,z_1)\overline{\psi_R}(y,z_2)\ket{0}=&-i\sum_nf_L^n(z_1)f_R^n(z_2)P_L\int \frac{d^4p}{(2\pi)^4}e^{ip(x-y)}\frac{m_n}{p^2+m_n^2-i\epsilon} \ , \\
  \bra{0}T \psi_R(x,z_1)\overline{\psi_L}(y,z_2)\ket{0}=&-i\sum_nf_R^n(z_1)f_L^n(z_2)P_R\int \frac{d^4p}{(2\pi)^4}e^{ip(x-y)}\frac{m_n}{p^2+m_n^2-i\epsilon} \ , \\
  \bra{0}T \psi_R(x,z_1)\overline{\psi_R}(y,z_2)\ket{0}=&-i\sum_nf_R^n(z_1)f_R^n(z_2)P_R\int \frac{d^4p}{(2\pi)^4}e^{ip(x-y)}\frac{-i\slashed p}{p^2+m_n^2-i\epsilon} \,
\eead
\ee
where $P_{L,R}$ are the left (right) projection operators. Using the action \eqref{5DactionCSUV} and computing Feynman diagrams such as those shown in Fig. \ref{CStriangle}, we can derive the one-loop effective Chern-Simons term that couples $V_M$ to the gluons,
\be
  \cL_{5D,CS}=-\frac{n_{CS}}{128\pi^2}\epsilon^{MNPQR}V_M
G_{NP}^cG_{QR}^c\ ,
\label{CStermShape}
\ee
where $n_{CS}$ is an integer that will be identified below.

For instance, the Feynman diagram that corresponds to the restriction of the Chern-Simons term to $V_z G^c \tilde G^c$ (in other words, the one corresponding to the scattering of the 4D scalar in $V_z$ and 4D gluons) reads
\be
  \delta^{(4)}(-p-q+p+q)4i\pi^2\epsilon^{\mu\nu\rho\sigma}p_\mu q_\rho\delta^{ab}\sum_{n=0}^{\infty}\frac{1}{m_n}\int dz A^4f_L^nf_R^n(z) \ .
\label{diagram5DVzgluons}
\ee
In order to match this expression to a local 5D Chern-Simons term, it is useful to notice, using the KK equations \eqref{KKeqs}, that
\be
  \partial_z\left(\sum_{n\geq 0}\frac{A^4}{m_n}f_L^n f_R^n\right)=A^4\sum_{n\geq 0}\left[(f_R^n)^2-(f_L^n)^2\right] \ .
\ee
The next step would be to use a completeness relation such as \eqref{completenessSumWithMixing}. For that, we need to specify our choice of boundary conditions for $\psi$, which determines the precise completeness relation satisfied by the KK profiles.

\begin{figure}[!h]
  \centering
  \includegraphics[width=0.4\textwidth]{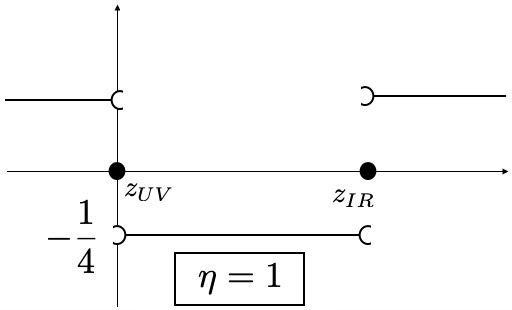}\quad\includegraphics[width=0.4\textwidth]{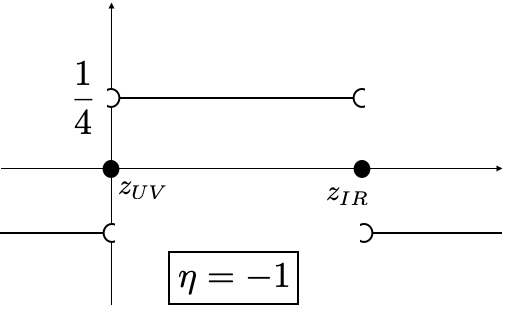}
  \caption{Plot of $\sum_{n\geq 0}\frac{A^4}{m_n}f_L^nf_R^n$ as a function of $z$, for opposite Dirichlet boundary conditions.}
  \label{plotSumKKOpposite}
\end{figure}

We will consider two cases. In the first case, we impose opposite Dirichlet boundary conditions on $\psi$, namely $(1-\eta\gamma_5)\psi(z_{UV})=0$ and $(1+\eta\gamma_5)\psi(z_{IR})=0$, with $\eta=\pm 1$. This means all KK modes are massive and can be integrated out. With such boundary conditions, \eqref{completenessSumWithMixing} becomes \cite{Hirayama:2003kk}
\be
  A^4\sum_n\left[(f_R^n)^2-(f_L^n)^2\right]=-\frac{\eta}{2}\left[\delta(z-z_{UV})-\delta(z-z_{IR})\right]\ .
\ee
Using this relation, we can deduce the form of $\sum_{n\geq 0}\frac{A^4}{m_n}f_L^nf_R^n$, which is plotted in Fig. \ref{plotSumKKOpposite}. 
Consequently, we can identify from \eqref{diagram5DVzgluons} the constant $n_{CS}$ in \eqref{CStermShape},
\be
  n_{CS}=-\eta \ .
\ee
In the second case, we impose Dirichlet boundary conditions only on $\psi_R$. There is now a left-handed zero mode which does not enter the Feynman diagrams, and therefore the sum in \eqref{diagram5DVzgluons} only runs over massive KK modes ($n>0$). This means the 5D fermion has not been completely integrated out, because there remains a left-handed zero mode in the low-energy spectrum. Consequently, \eqref{diagram5DVzgluons} cannot be generically matched to a local 5D Chern-Simons term. However, this conclusion can be avoided by taking the infinite bulk mass limit, $M_\psi\to\infty$. In non-compact 5D space, this is simply the limit where the bulk fermion decouples. Instead, on an orbifold, it corresponds to the limit where the KK modes decouple and the zero mode becomes localized on one of the branes. This implies that all effective bulk modes have been integrated out, and the result can be interpreted as a local term in 5D.

\begin{figure}[!h]
  \centering
  \includegraphics[width=0.4\textwidth]{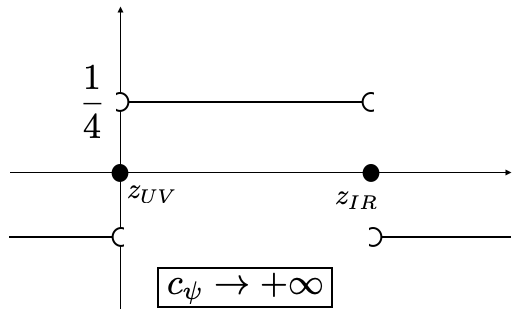}\quad\includegraphics[width=0.4\textwidth]{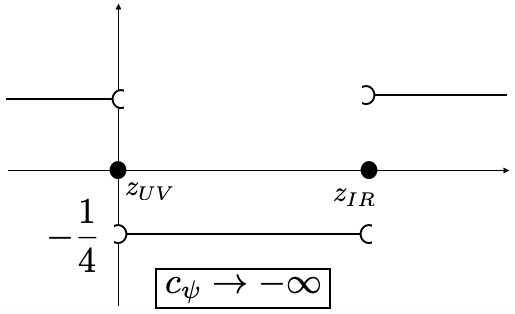}
  \caption{Plot of $\sum_{n>0}\frac{A^4}{m_n}f_L^nf_R^n$ as a function of $z$, for a left-handed zero mode.}
  \label{plotSumKK}
\end{figure}

Using again the KK equations \eqref{KKeqs} and the completeness relation \eqref{completenessSumWithMixing}, we obtain
\be
  \partial_z\left(\sum_{n>0}\frac{A^4}{m_n}f_L^n f_R^n\right)=A^4\sum_{n>0}\left[(f_R^n)^2-(f_L^n)^2\right]=-\frac{1}{2}\left(\delta(z-z_{IR})+\delta(z-z_{UV})\right)+A^4(f_L^0)^2 \ .
\label{eq:massiveKKsum}
\ee
The last term in \eqref{eq:massiveKKsum} prevents interpreting \eqref{diagram5DVzgluons} as arising from a local term in 5D, except if there is a limit where it becomes a Dirac delta function. This is precisely what happens in the infinite bulk mass limit. Indeed, using the explicit form of the zero mode profile in \eqref{eq:fermionprofiles}, we find
\be
  A^4(f_L^0(z))^2=\frac{(1-2c_\psi)k}{2\[(kz_{IR})^{1-2c_\psi}-(kz_{UV})^{1-2c_\psi}\]}(kz)^{-2c_\psi}\rightarrow \left\{\begin{matrix} \delta(z-z_{UV}), \qquad c_\psi\rightarrow +\infty \ \\ \delta(z-z_{IR}), \qquad c_\psi\rightarrow -\infty \  \end{matrix}\right. \ ,
\label{zeroModeLimit}
\ee 
with $M_\psi=c_\psi k$. Consequently, we obtain
\be
  \partial_z\left(\sum_{n>0}\frac{A^4}{m_n}f_L^nf_R^n\right)=
  \left\{\begin{matrix}  \frac{1}{2}\[\delta(z-z_{UV})-\delta(z-z_{IR})\], \qquad c_\psi\rightarrow +\infty \\  \frac{1}{2}\[\delta(z-z_{IR})-\delta(z-z_{UV})\], \qquad c_\psi\rightarrow -\infty \end{matrix}\right. \ ,
\ee
and, using the Dirichlet boundary conditions, we can deduce the form of $\sum_{n>0}\frac{A^4}{m_n}f_L^nf_R^n$, which is plotted in Fig. \ref{plotSumKK}.
Therefore, in the infinite bulk mass limit, the KK sum \eqref{diagram5DVzgluons} becomes local and the constant $n_{CS}$ in \eqref{CStermShape} can be identified as
\be
  n_{CS}=\text{sgn}(M_\psi) \ .
\ee
As known from \cite{Hirayama:2003kk}, the Chern-Simons term carries the full 5D anomaly for opposite Dirichlet boundary conditions. The final step to match our effective action to \eqref{eq:Chern-Simons} is to remove the IR anomaly. We discuss this in the next section. Instead, in the case of symmetric Dirichlet boundary conditions, adding the anomalous contribution of the Chern-Simons term to that of the boundary zero mode fermion leads to the 5D anomaly \eqref{eq:5Danomeqn} \cite{ArkaniHamed:2001is}. However, in order to match to the bosonic Lagrangian \eqref{eq:Chern-Simons}, it is useful to also give a mass to the 4D fermion zero mode and integrate it out. We also discuss this in the next section.

\subsection{Boundary axion terms}

We now introduce boundary terms to remove the IR anomaly, and obtain the effective action in \eqref{eq:Chern-Simons} for both choices of the boundary conditions.

When opposite Dirichlet boundary conditions are chosen, no fermionic degree of freedom remains in the IR, and all 5D anomalies are carried by the Chern-Simons term. To remove the IR anomaly, we can simply add the following boundary term
\be
  -\int d^4 x \frac{\eta}{64\pi^2}a G^c\tilde G^c\bigg\vert_{z_{IR}} \ .
\ee
This gives rise to \eqref{eq:Chern-Simons} with $\kappa=-\eta=\mp 1$ for left-right (right-left) fermion Dirichlet boundary conditions .

Instead, for symmetric boundary conditions, there remains a fermionic zero mode (that was chosen to be left-handed). A first set of boundary terms is intended to give a mass to the $\psi_L$ zero mode; focusing on the limit $M_\psi\to-\infty$, this is done by introducing a right-handed 4D fermion $\widetilde \psi_R$ on the IR brane and coupling it to $\psi_L$ via the following boundary term,
\be
  S_{5D}\supset -\int d^5 z\sqrt{-g}\frac{1}{\sqrt{g_{55}}}\delta(z-z_{IR})\(\frac{y}{(M_\psi k)^{1/2}}\overline\psi_L\widetilde\psi_R\Phi+h.c.\) \ ,
\ee
where $y$ is dimensionless coupling and the powers of $M_\psi,k$ are there to ensure the correct dimensions as well as a smooth $M_\psi\to\infty$ limit. In addition, the PQ and color charges of $\widetilde\psi_R$ are chosen appropriately. We can now integrate out the Dirac fermion $(\psi^{0}_L,\widetilde\psi_R)$ and we obtain the following axion coupling
\be
  -\int d^4 x \frac{1}{32\pi^2}a\, G^c\tilde G^c\bigg\vert_{z_{IR}} \ .
\label{axionontheIR}
\ee
Together with the Chern-Simons term, this boundary term ensures that the anomalous shift of the effective action has the form \eqref{eq:5Danomeqn}, as it should to reproduce the full 5D mixed anomaly between the PQ symmetry and QCD. Finally, to cancel the anomaly on the IR brane, we can simply add
\be
  \int d^4 x \frac{1}{64\pi^2}a G^c\tilde G^c\bigg\vert_{z_{IR}} \ .
\ee
Adding up the Chern-Simons term and all boundary terms, we obtain \eqref{eq:Chern-Simons} with $\kappa=-1$.


\bibliographystyle{JHEP}
\bibliography{flavoured-axion.bib}

\end{document}